\documentclass[12pt]{article}
\usepackage{rotating}
\usepackage{hhline}
\usepackage{array}
\usepackage{amsmath}
\usepackage{amssymb}
\usepackage{graphicx}
\usepackage{xcolor}
\usepackage{cite}
\usepackage{soul}
\usepackage{cancel}
\usepackage{comment}
\usepackage{hyperref}
\usepackage{cleveref}

\begin{document}

\tolerance=100000
\thispagestyle{empty}
\setcounter{page}{0}

\newcommand{\rb}[2]{\raisebox{#1}[-#1]{#2}}
\newcommand{\s}{\\ \vspace*{-3.5mm}}

\begin{titlepage}

\begin{flushright}
\today \\
KIAS-Q24009
\end{flushright}

\renewcommand{\thefootnote}{\fnsymbol{footnote}}

\mbox{ }
\vskip 2.cm

\begin{center}
{\Large \bf Vetoing all the Higgs imposters in 
   \boldmath{$H\to \ell^-\ell^+ Z$}} \\[1.cm]
   Seong Youl Choi$^a$\footnote{sychoi@jbnu.ac.kr},\,
   Jaehoon Jeong$^b$\footnote{jeong229@kias.re.kr}, and
   Dong Woo Kang$^a$\footnote{dongwookang@jbnu.ac.kr}\\[0.5cm]
{\it $^a$Laboratory for Symmetry and Structure of the Universe, 
           Department of Physics,\\
          Jeonbuk National University, Jeonju 54896, Korea\\
      $^b$School of Physics, Korea Institute for Advanced Study,
          Seoul 02455, Korea
}
\end{center}

\setcounter{footnote}{0}
\renewcommand{\thefootnote}{\arabic{footnote}}
\vspace{2cm}

\begin{abstract}
\noindent
We develop an effective and methodical algorithm for the construction 
of general covariant four-point $H\ell\ell Z$ vertices, accommodating 
leptons $\ell=e, \mu$, and designed to handle a boson $H$ of 
any integer spin, not merely confined to spins up to 2. While our 
numerical analysis assumes the $H$-boson mass to be 
$m_H=125\,{\rm GeV}$, the analytical framework we propose is versatile, 
enabling the examination of various mass as well as spin scenarios. 
These meticulously devised general covariant four-point $H\ell\ell Z$ 
vertices are pivotal in vetoing all the imposters of the Standard Model 
Higgs boson holding the spin-0 and even-parity quantum numbers,
especially in one of its primary decay channels, the three-body decay 
process $H\to \ell^-\ell^+ Z$, observable at the Large Hadron Collider.
Our innovative strategy encompasses the analysis of all the effectively 
allowed scenarios, extending beyond the limitations of previous 
investigations on the Higgs spin and parity determinations in the decay
$H \to \ell^-\ell^+ Z$. Based on the significantly expanded 
scheme, we demonstrate that the Higgs boson imposter of any spin and parity 
can be definitively vetoed by leveraging threshold effects and angular 
correlations, even though achieving such conclusive results in practical 
and exhaustive analyses necessitates high event rates.
\end{abstract}

\end{titlepage}

\setcounter{equation}{0}

\section{Introduction}
\label{sec:introduction}

The last missing piece of the beautiful tapestry of the Standard Model 
(SM)~\cite{Glashow:1961tr,Weinberg:1967tq,
Salam:1968rm, Fritzsch:1973pi} of particle physics has been filled by the
discovery~\cite{ATLAS:2012yve,CMS:2012qbp}
of a Higgs boson consistent 
with SM predictions at the Large Hadron Collider (LHC). As an
authoritative historical review on the SM, we refer to the 
work~\cite{Weinberg:2018apv} 
by Steven Weinberg, one of the SM founders.
Nevertheless, the theory of electroweak (EW) and strong interactions is 
strongly suggested 
to be a mere effective theory, underlying a more fundamental theoretical 
framework, due to various unresolved conceptual issues 
and unexplained experimental 
observations~\cite{ParticleDataGroup:2022pth}.\s

At present, in the absence of definitive and direct discoveries of new physics 
beyond the SM (BSM), we are confronting a revived period
characterized by precision-enhanced measurements and 
data-driven analyses. 
While the conventional model-building approach has 
significantly influenced the strategies for experimental measurements and designs 
in the quest for new physics over the past fifty years, we are now compelled 
to rely on a multitude of data-driven precise measurements. This approach is 
intended to facilitate the exploration of a more fundamental understanding 
of physics in a model-independent manner, both directly and indirectly. 
Recently, there has been increasing focus on these model-independent effective 
methods~\cite{Weinberg:2016kyd,Brivio:2017vri,Manohar:2018aog, 
Weinberg:2021exr,Burgess:2020tbq,Levi:2023gwz}, leading to
intensified efforts in developing tailored (data-driven) analytical 
approaches, as evidenced, for example, by the formation of 
the international seminar series called All Things 
EFT~\cite{all_things_eft}.\s

The development and evolution of the SM over more than five decades 
are anchored on two fundamental conceptual foundations. One pillar is the local gauge 
symmetry of the combined group SU(3)$_C\times$SU(2)$_L\times$U(1)$_Y$, governing the 
strong and unified electroweak interactions. The other pillar is the spontaneous 
breakdown of the EW symmetry SU(2)$_L\times$U(1)$_Y$ to the 
electromagnetic (EM) gauge symmetry U(1)$_{\rm EM}$, triggered by a non-zero vacuum 
expectation value (vev) of the neutral component of a single Higgs doublet field. 
This mechanism, first proposed in three seminal works 
in the year of 1964~\cite{Englert:1964et,Higgs:1964pj,
Guralnik:1964eu}, gives rise to a physical quantum fluctuation of the 
Higgs doublet field around the vacuum, manifesting 
as the massive SM Higgs boson announcing the occurrence of the 
spontaneous breaking of SM electroweak gauge symmetry. 
It holds specific spin $s_H$, parity $P$, and charge-conjugation 
$C$ quantum numbers:
\begin{eqnarray}
s_H[PC]\, =\, 0[++]\,.
\end{eqnarray}
These attributes are intrinsic to the 
interactions of the Higgs boson with EW gauge bosons, gluons, photons, and 
fermions, ensuring conservation of $C$, $P$, and $CP$ 
symmetries~\cite{Gunion:1989we,Djouadi:2005gi,
LHCHiggsCrossSectionWorkingGroup:2016ypw}. To solidify the foundations of 
the SM, it is crucial to ascertain the spin and parity of the discovered particle 
with more enhanced certainty, alongside precise measurements and 
determinations of its interactions with itself and all the SM 
particles across diverse production and decay channels 
at the 
LHC~\cite{Salam:2022izo,ATLAS:2022vkf,CMS:2022dwd,
ParticleDataGroup:2022pth,Ortona:2023ypp}
and upcoming lepton and 
hadron colliders~\cite{Mangano:2019kji,Shiltsev:2019rfl}.\s
 
Previous research on the systematic analysis of spin and parity quantum 
numbers has been extensively worked out in decays such as 
$Z^\star Z$~\cite{Choi:2002jk,Bredenstein:2006rh,
Bredenstein:2006nk,DeRujula:2010ys,Gao:2010qx,Bolognesi:2012mm,CMS:2012vby,
LHCHiggsCrossSectionWorkingGroup:2013rie,CMS:2013btf,ATLAS:2013xga,CMS:2013fjq,Choi:2013zqz,CMS:2014nkk,Beneke:2014sba,
Gainer:2014hha,CMS:2017len,Dawson:2018dcd}, 
$Z \gamma$~\cite{Stolarski:2012ps,Low:2012rj,Corbett:2012dm,Anderson:2013afp}, 
and $\gamma\gamma$ decays~\cite{Ellis:2012wg,Ellis:2012jv,Alves:2012fb,
Choi:2012yg,Davis:2021tiv},
and $CP$-violating decays~\cite{Soni:1993jc,Chang:1993jy,
Godbole:2007cn,Nelson:1989uq,Grzadkowski:1995rx}. 
These analyses have explored diverse 
production channels like gluon and vector boson 
fusion~\cite{Plehn:2001nj,Hagiwara:2009wt,Englert:2012xt}, 
Higgs-strahlung and 
associated Higgs-vector production~\cite{Barger:1993wt,
Miller:2001bi,Ellis:2013ywa}.
Nevertheless, additional challenges persist in accurately establishing the spin 
and parity of the SM Higgs boson~\cite{Artoisenet:2013puc}. 
One crucial challenge arises due to limitations in the theoretical
frameworks used in previous analyses, which do not 
comprehensively cover all potential scenarios.

\begin{figure}[ht!]
\vskip 0.1cm
\centering
\includegraphics[scale=0.35]{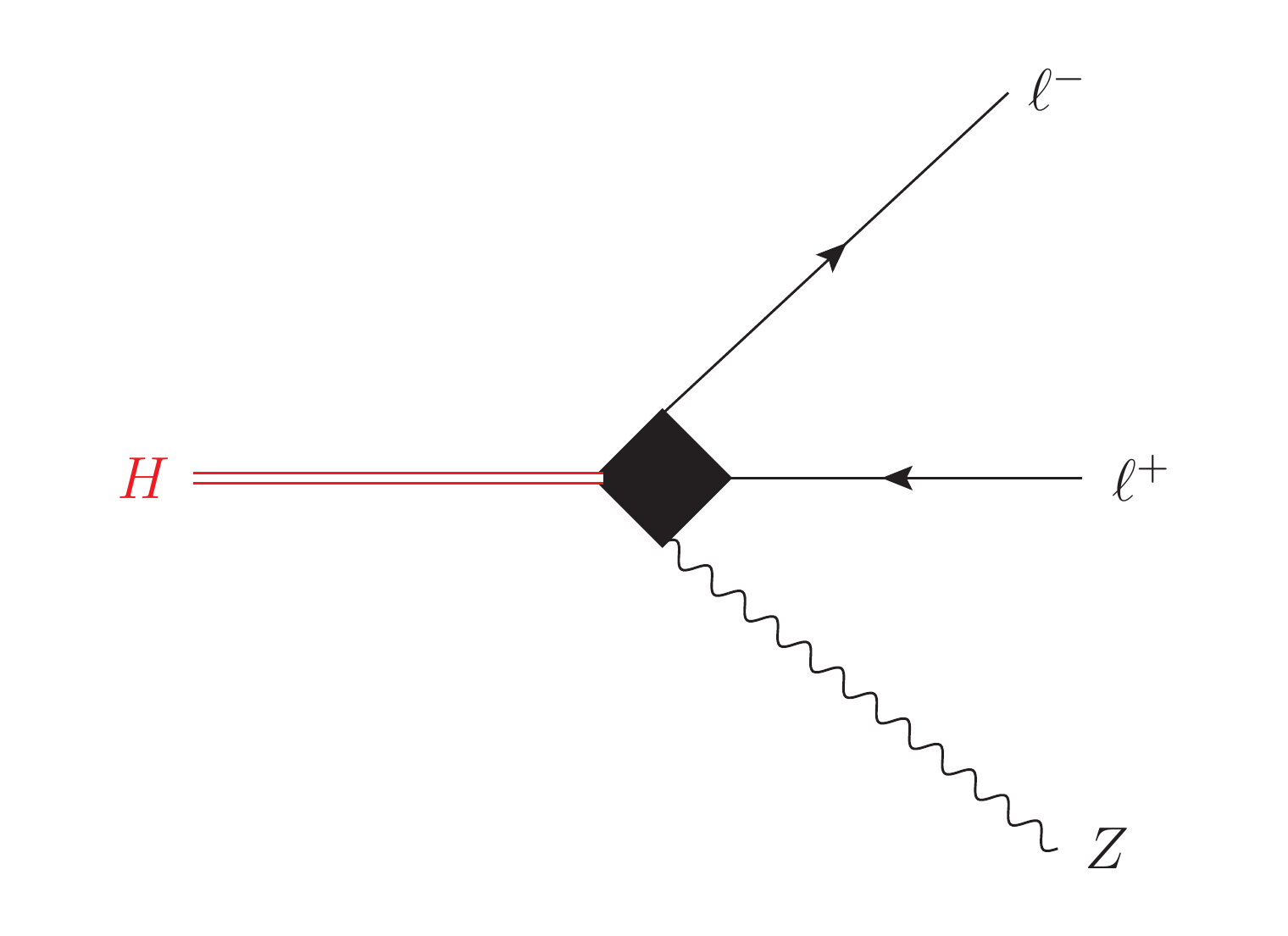}
\caption{\it Diagrammatic description of the four-point $H\ell\ell Z$ 
            vertex the decay process $H\to \ell^-\ell^+ Z$ for
             a massive $H$ of any integer spin. In the SM,
             the particle $H$ can be identified to be the SM Higgs boson
             with its definite discrete spacetime quantum numbers and 
             well-measured mass $m_H=125\, {\rm GeV}$.
             The lepton $\ell$ is assumed to be $e$ or $\mu$ with its
             mass ignored.
             }
\label{fig:4-point_vertex_diagram}
\end{figure}

In particular, one part of  the past analyses that aimed to identify the spin 
and parity of the SM Higgs boson via the 3-body decay $H\to\ell^-\ell^+ Z$ 
with $\ell=e, \mu$ have relied on a two-step cascade process involving an 
intermediate virtual $Z^\star$ exchange, followed by 
a leptonic SM $Z$ decay $Z^\star\to \ell^-\ell^+$, in the consideration 
of the $H$ boson of any integer spin~\cite{Choi:2002jk,Bredenstein:2006rh,
Bredenstein:2006nk,DeRujula:2010ys,Gao:2010qx,Bolognesi:2012mm,
LHCHiggsCrossSectionWorkingGroup:2013rie,CMS:2013btf,
CMS:2012vby,ATLAS:2013xga,CMS:2013fjq,CMS:2014nkk,Choi:2013zqz,Beneke:2014sba,
Gainer:2014hha,CMS:2017len,Dawson:2018dcd,Stolarski:2012ps,Low:2012rj,
Corbett:2012dm,Anderson:2013afp}. 
The other part of them has been worked out by considering the process 
$H\to \ell^-\ell^+ Z$ occurring via the contact as well as all 
cascade-channel interactions mainly for the spinless 
$H$ boson~\cite{Rao:2006hn,Rao:2007ce,Huitu:2015rha,
Grinstein:2013vsa,Gritsan:2020pib}.
In order to ensure a complete and unambiguous determination of 
the Higgs spin and parity, it is essential to consider a significantly broader 
and more general scope, including the general 4-point $H\ell\ell Z$ 
vertex depicted with a black square in Fig.$\,$\ref{fig:4-point_vertex_diagram} 
that represents all potential interactions, encompassing not only the specific 
two-step cascade process but also additional scenarios involving all 
permissible channel processes and four-point contact interactions among 
the four particles, $H$, $\ell^\mp$, and $Z$, with the $H$ boson of 
any integer spin. Building upon this substantially expanded framework, 
we undertake a model-independent and thorough analysis to eliminate 
all potential Higgs imposters via the decay $H\to\ell^-\ell^+ Z$, 
where $\ell=e, \mu$. This approach contributes to a more refined 
determination of the spin and parity properties of the Standard Model 
Higgs boson.\s 

As the initial step, we develop an efficient and systematic 
algorithm for constructing general covariant four-point $H\ell\ell Z$ 
vertices with  $\ell=e, \mu$ for a particle $H$ of any integer spin of which
the value does not have to be restricted up to two. 
Although the $H$ mass is numerically set to 
$m_H=125\, {\rm GeV}$~\cite{ParticleDataGroup:2022pth},
our method allows for flexibility in handling different mass values. 
These carefully constructed general covariant four-point $H\ell\ell Z$ 
vertices play a crucial role in systematically identifying the spin 
and parity of the SM Higgs boson through its prime Higgs
decay channel, the three-body process $H\to \ell^-\ell^+ Z$, where 
$\ell=e, \mu$, at the LHC. We emphasize again that this novel approach 
surpasses previous analytic platforms for simulation and 
experimental studies on Higgs spin and parity 
determination~\cite{CMS:2012qbp, Choi:2002jk,Bredenstein:2006rh,
Bredenstein:2006nk,DeRujula:2010ys,Gao:2010qx,Bolognesi:2012mm,
Stolarski:2012ps,LHCHiggsCrossSectionWorkingGroup:2013rie,CMS:2013btf,
CMS:2012vby,ATLAS:2013xga,CMS:2013fjq,Choi:2013zqz,Beneke:2014sba,
Gainer:2014hha,CMS:2017len,Dawson:2018dcd}.\s

\begin{figure}[ht!]
\vskip 0.4cm
\centering
\includegraphics[scale=1]{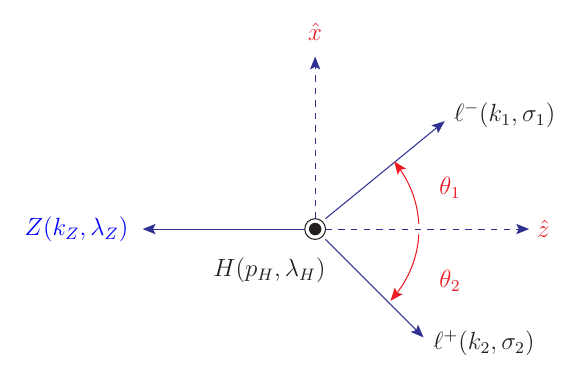}
\caption{\it Kinematic configuration of the three-body decay
            $H(p_H,\lambda_H) \to \ell^-(k_1, \sigma_1)
             + \ell^+(k_2,\sigma_2)+ Z (k_Z, \lambda_Z)$ in
             the rest frame of the $H$ boson. Here, $p_H$, $k_1$, $k_2$ and
             $k_Z$ are the $H$, $\ell^-$, $\ell^+$ and $Z$ four-momenta,
             while $\lambda_H$, $\sigma_1$, $\sigma_2$ and $\lambda_Z$ are
             their helicities, respectively. $\theta_1$ and $\theta_2$
             are the polar angles of the $\ell^-$ and $\ell^+$ momentum 
             directions with respect to the direction opposite to the
             $Z$ momentum direction, which is set to be the positive
             $z$-axis. The decay plane is defined to be the $xz$ plane
             without any loss of generality, as it will become clear 
             after taking into account the on-shell $Z$ boson decay
             into a fermion-antifermion pair.  
             }
\label{fig:h_to_llz_rm}
\end{figure}

By leveraging the one-to-one correspondence between the helicity 
formalism and the covariant formulation, we construct the general 
covariant four-point $H\ell\ell Z$ vertices to describe the 
3-body decay of the $H$ boson with any integer spin $s_H$
\begin{eqnarray}
  H(s_H,m_H)
\,\,\rightarrow\,\,
   \ell^-(1/2,m_\ell)
   +\ell^+(1/2,m_\ell)
   + Z(1,m_Z),
\label{eq:ee_ZX_scattering_process}
\end{eqnarray}
in the $H$ rest frame where $\{s_H,1/2,1\}$ and $\{m_H,m_\ell,m_Z\}$ are 
the spins and masses of the particles, $\{H,\ell^\mp,Z\}$, respectively.
The methodology for constructing these vertices is a logical 
and challenging extension of the algorithm used in prior works to build 
general three-point vertices, as detailed in a sequence of 
publications~\cite{Choi:2021ewa,Choi:2021qsb,Choi:2021szj}.
The kinematic configuration of the 3-body decay process in the $H$ rest frame 
is depicted in Fig.~\ref{fig:h_to_llz_rm},
where $p_H$, $k_1$, $k_2$ and $k_Z$ denote the four-momenta for the $H$ boson,
two leptons $\ell^\mp$ and the $Z$ boson, respectively, and 
$\lambda_H$, $\sigma_{1,2}$ and $\lambda_Z$ denote the $H$, 
$\ell^\mp$ and $Z$ helicities, respectively. The momentum direction of 
the $\ell^-\ell^+$ system is 
taken to be along the positive $z$-axis, while the $Z$-boson momentum direction 
is set to be along the negative z-axis.

Including the leptonic decays
of the on-shell $Z$ boson $Z\to f\bar{f}$ with $f=e$ and $\mu$, 
the clean final 4-lepton channels allow us to isolate the signal
of the 3-body $H$ decay from the background effectively and 
to construct the kinematical 
configuration completely with good precision. While the dominant decay
mode for the Higgs mass of 125 GeV is the $b\bar{b}$ channel, the
$Z^\star Z$ mode with one virtual $Z^\star$ boson below the threshold
for two real $Z$ bosons is one of the sub-leading channels next 
to the $W^\star W$ channel~\cite{Djouadi:1997yw,Djouadi:2018xqq}. 
We show that the $\ell^-\ell^+$ invariant-mass distributions and 
various angular correlations in the 3-body decay of the $H$ boson 
allow for a clear determination of the spin and parity of the SM Higgs 
boson. This determination remains feasible in such a significantly enlarged 
scenario that encompasses the general $H\ell\ell Z$ interaction with 
the $H$ boson with any integer spin $s_H$, although conducting practical 
and comprehensive analyses require high event rates.\s

The paper is organized as follows. In Sec.$\,$\ref{sec:wave_functions}, 
we work out the kinematics of the on-shell $Z$ boson and two 
charged leptons $\ell^\mp$  in detail, derive two chirality-conserving 
(CC) $\ell\ell$ vector currents in the massless limit with $m_{\ell}=0$ 
and introduce the general form of integer-spin wave functions expressed as a linear 
combination of products of spin-1 polarization vectors coupled with 
the appropriate Clebsch-Gordon coefficients in the $H$ rest frame. 
All of them play an essential role in constructing the general 
structure of the covariant 4-point $H\ell\ell Z$ vertices.
In Sec.~\ref{sec:constructing_helicity-specific_operators} 
we define and construct all the basic $HZ$ and $\ell\ell$ 
helicity-specific operators by investigating  the decays of a spin-1 
particle into two spin-1 massive particles and two spin-1/2 massless 
particles, respectively.
Following the construction of the five fundamental $HZ$ operators and 
the two essential $\ell\ell$ operators that give rise to the CC lepton 
vector currents in Sec.~\ref{sec:constructing_helicity-specific_operators},
we meticulously outline an effective and methodical algorithm for constructing 
all the helicity-specific covariant $H\ell\ell Z$ vertices in 
Sec.$\,$\ref{sec:weaving_covariant_4-point_vertices}. 
This development stands as the pivotal outcome of our current study.
Section~\ref{sec:sm_higgs_decays} describes
all the key characteristics of the invariant mass distribution and
angular correlations specific to the SM Higgs boson. After 
describing the characteristic features of the SM Higgs boson,
we demonstrate 
in Sec.$\,$ \ref{sec:vetoing_higgs_imposters} that any scenario 
of a $H$ imposter with any integer spin or/and odd parity can be ruled 
out unambiguously through the effective cooperation of the invariant mass  
distribution and angular correlations.  We emphasize that unlike most 
previous studies, our comprehensive studies include all the allowed 
4-point $H\ell\ell Z$ processes, in addition to the standard two-step
processes, $H\to Z^\star Z\to \ell^-\ell^+ Z$. 
Section~\ref{sec:summary_conclusion} summarizes several key findings 
and concludes with a definitive and persuasive statement on a more 
refined identification of the spin and parity of the Standard Model Higgs 
through its primary $H$ decay mode.
In addition, several 
analytic formulas to be utilized for evaluating all the helicity amplitudes 
are collected in Appendix~\ref{appendix:analytic_formulas}
and  a set of integral functions to be used for
deriving all the invariant-mass and polar- and azimuthal-angle correlations 
are listed in Appendix~\ref{appedix:integral_functions}.\s

\setcounter{equation}{0}

\section{Kinematics and wave functions}
\label{sec:wave_functions}

To systematically develop our algorithm for constructing general 
covariant four-point $H\ell\ell Z$ vertices and calculating all relevant decay 
amplitudes and correlations, it is imperative to utilize various kinematic
quantities and wave functions. These elements are fundamental in determining
the vertex structure. Section~\ref{sec:wave_functions} 
is devoted to listing them collectively.\s

As the electron and muon masses, $m_e=5.1\times 10^{-3} \, {\rm GeV}$ and
$m_\mu=1.06\times 10^{-1}\, {\rm GeV}$, are significantly
smaller than the scale $v=246\,{\rm GeV}$ of spontaneous electroweak
symmetry breaking (EWSB)~\cite{ParticleDataGroup:2022pth}, 
it is justifiable and generally safe to disregard the masses of light leptons 
and concentrate solely on the CC interactions from a conceptual standpoint 
and practical perspective. This approach is valid since the 
chirality-violating (CV) couplings typically scale with the tiny
lepton masses.\footnote{The couplings of the $H$ boson to light
leptons and all other effects proportional to $m_{e, \mu}/m_H$ are
ignored.} Then, the $\ell^-\ell^+$ bi-linear term for
$\ell=e, \mu$  is assumed to be given by the CC
right/left-handed vector current defined by\footnote{It is straightforward to 
include the CV $\ell^-\ell^+$ scalar currents, although not considered in the 
present work.}
\begin{eqnarray}
   L^\mu_\pm
=  \bar{u}(k_1,\pm\mbox{$\frac{1}{2}$}) \,
   \gamma^\mu P_\pm \,
   v(k_2,\mp \mbox{$\frac{1}{2}$})
= -\sqrt{2} i m_\star \,\hat{l}_\pm \quad
  \mbox{with}\quad
  P_\pm = \frac{1\pm \gamma_5}{2},
\label{eq:chiral_ll_vector_current}
\end{eqnarray}
with $k=k_1+k_2$ and  $m_\star=\sqrt{k^2}$ is the $\ell^-\ell^+$ invariant 
mass. Here, the normalized space-like four vectors $\hat{l}_\pm$
satisfy the following orthonormal relations in the limit with $m_\ell = 0$
\begin{eqnarray}
   \hat{l}_+ \cdot{\hat{l}}^{*}_+
 = \hat{l}_- \cdot{\hat{l}}^{*}_-
 = -1,\qquad
   \hat{l}_+\cdot\hat{l}^*_- 
 = \hat{l}_-\cdot\hat{l}^*_+
 = 0,
\end{eqnarray}
valid in any reference frame. \s

For the sake of notation, we introduce the following dimensionless kinematic 
quantities 
\begin{eqnarray}
\omega_{\star, Z}=m_{\star, Z}/m_H,
 \quad
\eta^{\pm}=\sqrt{1-(\omega_\star\pm\omega_Z)^2},
 \quad
\kappa=\eta^+\eta^-,
 \quad
e_{\star, Z}=1\pm(\omega^2_\star-\omega^2_Z),
\end{eqnarray}
which enable us to parameterize the $H$, $\ell^-$, $\ell^+$, and $Z$ 
four-momenta as
\begin{eqnarray}
&&  
k=k_1+k_2=p_H-k_Z = m_\star\,\hat{k},\quad\ \
    l=k_1-k_2 = m_\star\, \hat{l},
\label{eq:kl_reparameterization} \\[2mm]
&&   p_H =k+k_Z= m_H\, \hat{p},\quad\ \
    q=k-k_Z
   =m_H\,(\omega_\star^2-\omega_Z^2)\,\hat{p}
    +m_H\,\kappa \,\hat{r},
\label{eq:4-momenta_reparameterization}
\end{eqnarray}
in terms of four normalized momenta, $\hat{p}$, $\hat{k}$, $\hat{l}$ and $\hat{r}$,
satisfying the normalization  conditions, 
$\hat{p}^2=\hat{k}^2=1$ and $\hat{l}^2=\hat{r}^2=-1$.
The normalized momenta, $\hat{p}$ and $\hat{r}$, are simply 
given by
\begin{eqnarray}
\hat{p} = (1,\, 0,\, 0,\, 0),\quad
\hat{r} = (0,\, 0,\, 0,\, 1),
\label{eq:hat_p_and_hat_r}
\end{eqnarray}
in the $H$ rest frame with the kinematic
configuration in Fig.$\,$\ref{fig:h_to_llz_rm}.
The normalized momenta, $\hat{k}$ and $\hat{l}$,
and two polar angles, $\theta_1$ and $\theta_2$, are given by
\begin{eqnarray}
&& \hat{k} = (\gamma_\star, 0, 0, \gamma_\star\beta_\star), \qquad\ \
   \hat{l} = (\gamma_\star\beta_\star\cos\theta, \sin\theta, 0,
              \gamma_\star\cos\theta),\\[2mm]
&& \cos\theta_1 = \frac{\beta_\star+\cos\theta}{1+\beta_\star\cos\theta},
   \quad\ \
   \cos\theta_2 = \frac{\beta_\star-\cos\theta}{1-\beta_\star\cos\theta},
\end{eqnarray}
and the space-like normalized vectors $\hat{l}_\pm$ in Eq.~\eqref{eq:chiral_ll_vector_current}, are orthogonal to both
$\hat{k}$ and $\hat{l}$, and written as
\begin{eqnarray}
  \hat{l}_\pm 
=\pm \frac{1}{\sqrt{2}}\, (\gamma_\star \beta_\star\sin\theta, 
                          -\cos\theta, \pm i, \gamma_\star\sin\theta),
\end{eqnarray}
in terms of the angle $\theta$ denoting the polar angle of 
the lepton $\ell^-$  with respect to the positive $z$ axis in the rest frame 
of the $\ell^-\ell^+$ system. The two boost parameters, $\gamma_\star$ and
$\beta_\star$, describing the Lorentz transformation of the $\ell^-\ell^+$ 
system from the CM frame to the $H$ rest frame and the other two boost 
parameters, $\gamma_Z$ and $\beta_Z$, describing that of the on-shell $Z$ boson 
from the rest frame to the $H$ rest frame are
\begin{eqnarray}
\gamma_\star = \frac{e_\star}{2\omega_\star},\quad
\beta_\star  = \frac{\kappa}{e_\star};\quad 
\gamma_Z     = \frac{e_Z}{2\omega_Z},\quad
\beta_Z      = \frac{\kappa}{e_Z},
\label{eq:boost_factors}
\end{eqnarray}
in the coordinate system defined in 
Fig.$\,$\ref{fig:h_to_llz_rm}.
It is noteworthy that the normalized momenta, 
$\hat{p}$ and $\hat{k}$, are time-like but the normalized momenta 
$\hat{r}$, $\hat{l}$, and $\hat{l}_\pm$, are space-like. The 
invariant products, $\hat{p}\cdot\hat{r}=\hat{k}\cdot\hat{l}=0$, 
are vanishing in any reference frame, while the invariant products, 
$\hat{l}\cdot\hat{p}$ and $\hat{l}\cdot\hat{r}$ are 
$\gamma_*\beta_\star\cos\theta$ and $\gamma_*\cos\theta$, respectively, 
in the $\ell^-\ell^+$ CM frame.\s

The wave function of a spin-0 particle is simply $1$, which is independent 
of its mass and momentum.
On the other hand, for a non-zero integer spin $s$, the wave function
of an incoming massive boson with its momentum $p$ and helicity
$\lambda$ is given by a wave tensor defined as a linear combination
of products of $s$ spin-1 polarization vectors with appropriate Clebsh-Gordon
coefficients by~\cite{Behrends:1957rup,Auvil:1966eao,
Caudrey:1968vih,Scadron:1968zz,Chung:1997jn,Huang:2003ym}
\begin{eqnarray}
   \epsilon^{\mu_1\cdots \mu_s}(p,\lambda)
=  \sqrt{\frac{2^s(s+\lambda)!(s-\lambda)!}{(2s)!}}
   \sum_{\{\tau_i\}=\pm 1, 0} \delta_{\tau_1+\cdots +\tau_s,\,\lambda}\,
   \prod^s_{j=1}
   \frac{\epsilon^{\mu_j}(k,\tau_j)}{\sqrt{2}^{|\tau_j|}}\, ,
\label{eq:bosonic_wave_tensor}
\end{eqnarray}
with the abbreviation $\{\tau_i\}=\tau_1,\cdots, \tau_s$.
The totally symmetric, traceless, and divergence-free wave function 
\eqref{eq:bosonic_wave_tensor} enables us to construct all the related 
interaction vertices efficiently as will be demonstrated later.
We note in passing that if the integer-spin particle is
massless, the wave tensor has only two maximal-magnitude helicities
of $\lambda=\pm s$ and its form is given by a direct product
of $s$ spin-1 wave vectors, each of which carries the 
same helicity of $\pm 1$.
The transverse and longitudinal-polarization vectors of a massive
spin-1 particle read
\begin{eqnarray}
  \epsilon(p,\pm 1)
= \frac{1}{\sqrt{2}}\, 
  (0,\, \mp \hat{\theta}-i \hat{\phi}),\ \
 \epsilon(p, 0)
= \frac{1}{m} (|\vec{p}\,|,\, E\hat{n}),
\label{eq:general_polarization_vectors}
\end{eqnarray}
in the helicity basis with the polarization axis along the 
normalized momentum direction 
$\hat{n}=(\sin\theta \cos\phi, \sin\theta \sin\phi,\cos\theta)$,
forming an orthonormal basis together with two unit vectors 
$\hat{\theta}=(\cos\theta\cos\phi,\cos\theta\sin\phi, -\sin\theta)$
and $\hat{\phi}=(-\sin\phi, \cos\phi, 0)$.\s

In the kinematic configuration of Fig.$\,$\ref{fig:h_to_llz_rm} 
where the $H$ boson is at rest with its 4-momentum $p_H=(1,0,0,0)$, 
the polarization axis 
of the $H$ boson can be freely chosen so that it will be set to be the 
positive $z$-axis for any explicit analytic evaluation in the following.
Then, the incoming polarization vectors of the $H$ boson are given by
\begin{eqnarray}
\epsilon_H(p_H,\pm 1)
  = \frac{1}{\sqrt{2}}\,
     (0,\, \mp 1, -i, 0),\qquad
    \epsilon_H(p_H, 0)
  = (0, 0, 0, 1).
\label{eq:h_polarization_vector}
\end{eqnarray}
On the other hand, the $Z$ polarization vectors in its helicity
basis with the polarization axis along the negative $z$-axis are 
written as
\begin{eqnarray}
 \epsilon_Z(k_Z,\pm 1)
  = \frac{1}{\sqrt{2}}\,
     (0,\, \mp 1, i, 0),\qquad\,
   \epsilon_Z(k_Z, 0)
  = \frac{1}{2\,\omega_Z}
     (\kappa,\, 0, 0, -e_Z),
\end{eqnarray}
in the helicity formalism with the Wick 
convention~\cite{Wick:1962zz,Dreiner:2008tw,Dreiner:2023yus,
Martin:1970hmp,Leader:2011vwq, Choi:2019aig} slightly
modified from the so-called Jacob-Wick (JW) 
convention~\cite{Jacob:1959at}.\s

\setcounter{equation}{0}

\section{Constructing helicity-specific operators}
\label{sec:constructing_helicity-specific_operators}

In this section, we define and construct all the basic $HZ$ and 
$\ell\ell$ helicity-specific operators 
necessary for weaving the general 4-point $H\ell\ell Z$ vertices 
systematically and compactly.\s

The initial step is to derive all five Lorentz-covariant basic $HZ$ 
helicity-specific operators by investigating the two-body decay 
$H \rightarrow XZ$  of a spin-1 massive particle $H$ into two spin-1 
massive particles $X$ and $Z$. Each of these operators generates a nonzero 
decay amplitude solely for a specific configuration $[\lambda_H,\lambda_Z]$ 
of the $H$ and $Z$ helicities in the $H$ rest frame when coupled to 
the wave functions~\cite{Choi:2021ewa,Choi:2021qsb,Choi:2021szj}.
Their expressions and roles in transitions originating from the $[0,0]$ 
point in the $HZ$ helicity space are articulated as follows:
\begin{eqnarray}
     U^0_{\beta\alpha}\hat{p}_{\star\mu}
 &=& \hat{p}_{Z\beta}
 \hat{r}_{H\alpha}   \hat{p}_{\star\mu}
    \qquad\qquad\qquad\quad\quad
    \!\!\!\,\rightarrow
    \quad
    [\lambda_H,\lambda_Z]=[0,0],
\label{eq:U00_al_be_basic_opertor} \\
U^\pm_{\mu\alpha}\hat{p}_{Z\beta}
 &=& \frac{1}{2} \left[g_{\bot\mu\alpha} \pm i
    \langle  \mu\alpha\hat{p}\hat{r}\rangle\right]
    \hat{p}_{Z\beta}
    \quad\quad\ \ \!\!
    \rightarrow
    \quad[\lambda_H,\lambda_Z]=[\pm 1,0],
\label{eq:U+-_al_mu_basic_opertor} \\
U^\pm_{\mu\beta} \hat{r}_{H\alpha}
&=&  \frac{1}{2} \left[g_{\bot\mu\beta} \pm i
    \langle\mu\beta\hat{p}\hat{r}\rangle\right]
    \hat{r}_{H\alpha}
    \quad\ \ \,\,\,
    \rightarrow
    \quad
   [\lambda_H,\lambda_Z] =[0,\pm 1].
\label{eq:U+-_be_mu_basic_opertor}
\end{eqnarray}
where the four-vector indices, $\alpha$, $\mu$, and $\beta$, are 
associated with the spin-1 $H$, $X$, and $Z$ bosons, respectively.
These are defined in terms of two normalized momenta, $\hat{p}$ and 
$\hat{r}$, in Eq.~\eqref{eq:4-momenta_reparameterization}, and 
three re-scaled momenta: $\hat{r}_{H\alpha}=\kappa\hat{r}_\alpha$, 
$\hat{p}_{\star\mu}= 2\omega_\star\hat{p}_\mu$, and
$\hat{p}_{Z\beta}=2\omega_Z\hat{p}_\beta$. 
The re-scaled momentum $\hat{r}_{H\alpha}=\kappa \hat{r}_\alpha$ is 
intentionally introduced because the normalized space-like four-vector $\hat{r}$
is always accompanied with the kinematic factor $\kappa$
as shown in Eq.$\,$(\ref{eq:4-momenta_reparameterization}).\s

\begin{figure}[ht!]
\vskip 0.5cm
\centering
\includegraphics[scale=0.33]{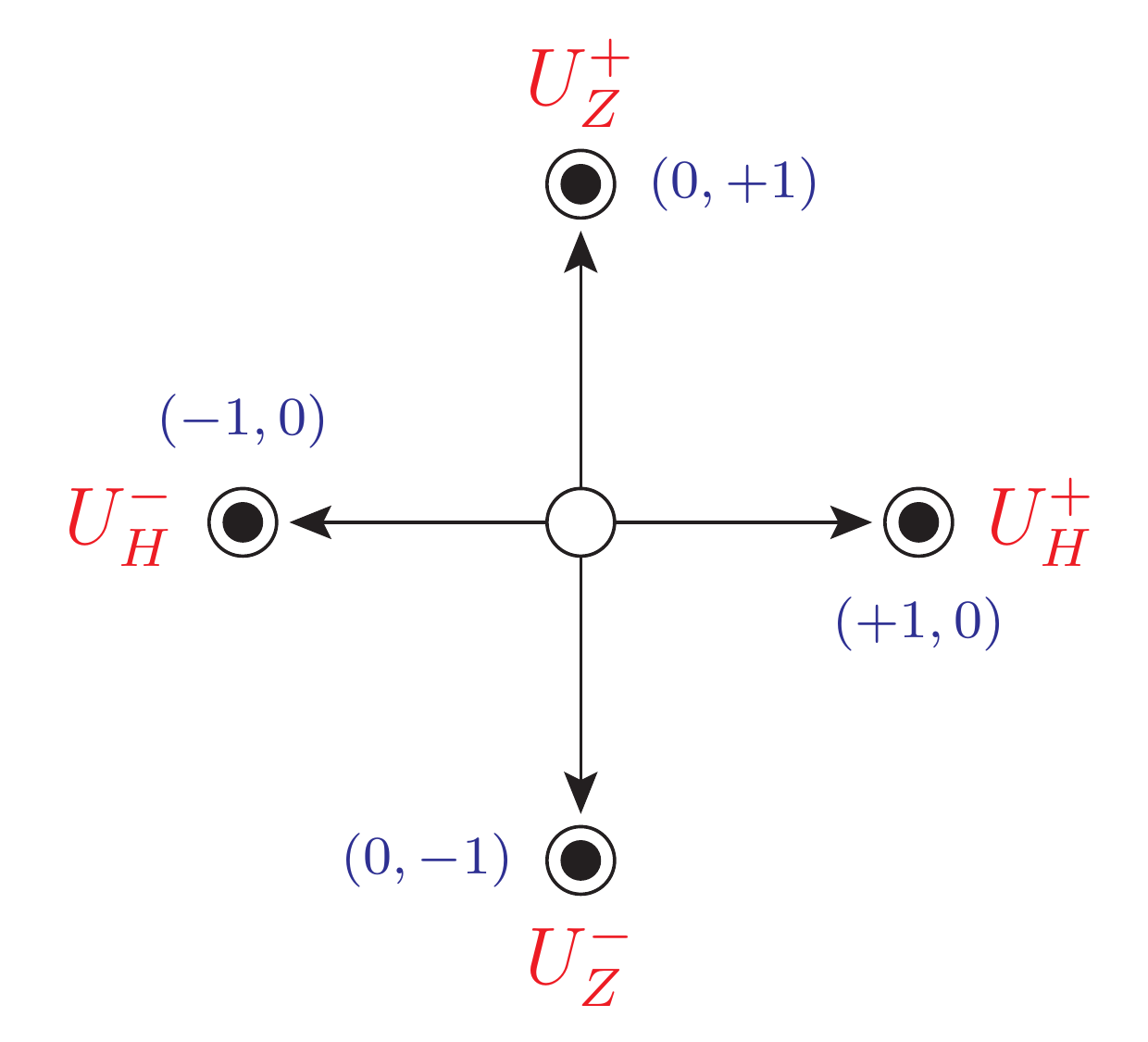}
\caption{\it Diagrammatic description of the four bosonic
   $U^\pm_H$ and $U^\pm_Z$ operators. In components,
   $[U^\pm_H]_{\alpha\mu}=U^\pm_{\alpha\mu}$ and
   $[U^\pm_Z]_{\beta\mu}=U^\pm_{\beta\mu}$, which are
   defined simply by the same set of two basic operators $U^\pm$.
   The detailed expressions are listed in
   Eqs.$\,$(\ref{eq:U+-_al_mu_basic_opertor})
   and (\ref{eq:U+-_be_mu_basic_opertor}).
 }
\label{fig:four_basic_bosonic_operators}
\end{figure}

For conceptual clarity and compact notation, 
we introduce  the square-bracket notations as
$U^\pm_{\alpha\mu}=[U^\pm_H]_{\alpha\mu}$, $U^\mp_{\beta\mu}=[U^\pm_Z]_{\beta\mu}$ 
and $U^0_{\alpha\beta}=[U^0_{HZ}]_{\alpha\beta}
=\hat{r}_{H\alpha} \hat{p}_{Z\beta}$.
The implications of the four basic 
$HZ$ operators, $U^\pm_H$ and $U^\pm_Z$, governing 
the horizontal and vertical one-step helicity transitions in 
the $HZ$ helicity space can be illustrated in 
Fig.$\,$\ref{fig:four_basic_bosonic_operators}.
The two orthogonal tensors $g_{\bot \mu\nu}$ and 
$\langle \mu\nu\hat{p}\hat{r}\rangle$ are defined by
\begin{eqnarray}
    g_{\bot \mu\nu}
&=& g_{\mu\nu}-\hat{p}_\mu \hat{p}_\nu + \hat{r}_\mu \hat{r}_\nu,\\
   \langle \mu\nu\hat{p}\hat{r}\rangle
&=& \varepsilon_{\mu\nu\rho\sigma}\, \hat{p}^\rho \hat{r}^\sigma,
\end{eqnarray}
in terms of the totally antisymmetric Levi-Civita tensor 
$\varepsilon_{\mu\nu\rho\sigma}$
with the convention of $\varepsilon_{0123}=+1$.
Contracting two basic $HZ$ operators, $U_H^\pm$ and $U_Z^\mp$, lead
to two composite transition operators governing the right-down 
and left-up diagonal one-step transitions in the $HZ$ helicity space as
\begin{eqnarray}
 U^{\pm}_{\beta\alpha}
= g^{\mu\nu} U^{\mp}_{\mu\beta} U^{\pm}_{\nu\alpha}
= 
\frac{1}{2} \left[g_{\bot\beta\alpha} \pm i
    \langle  \beta\alpha\hat{p}\hat{r}\rangle\right]
    \quad\ \
    \rightarrow
    \quad[\lambda_H,\lambda_Z]=[\pm 1,\mp 1],\,
\end{eqnarray}
each of which yields a nonzero decay amplitude only for the configuration 
$[\lambda_H,\lambda_Z]=[+1,- 1]$ or $[-1,+1]$ of the $H$ and $Z$ helicities.\s

In the second phase of our approach, we introduce the covariant CC 
basic $\ell\ell$ operators $W_\pm$ re-scaled by the $H$ mass
$m_H$, leading to the dimensionless amplitude of a two-body decay 
$X \rightarrow \ell^-\ell^+$ for a spin-1 particle $X$:
\begin{eqnarray}
    W^{\mu}_{\pm}
=  \frac{i}{\sqrt{2} m_H}\gamma^\mu P_\pm ,
\end{eqnarray}
with the chiral projection operators $P_\pm = (1\pm \gamma_5)/2$. 
The imaginary number
$i=\sqrt{-1}$ is factored out for removing it emerging in the helicity
amplitude computation according to the Wick convention.

In the third step, given the complexity arising from the involvement of many 
4-vector indices, especially in the context of high-spin states, 
we adopt a set of compact square-bracket notations. These notations are 
designed to succinctly represent the diverse array of indices encountered 
in the expression of general covariant vertices. Accompanying these notations, 
we also provide their respective component expressions to ensure clarity and 
facilitate the understanding of their roles within the general vertices 
as 
\begin{eqnarray}
&&  [\;\hat{p}_\star\;]
   \qquad\ \ \rightarrow \qquad
  \hat{p}_{\star \mu} = 2\omega_*\,\hat{p}_{\mu},
\\
&& [\, \hat{p}_Z\, ]
   \qquad\ \ \rightarrow \qquad
  \hat{p}_{Z \beta} = 2\omega_Z\,\hat{p}_{\beta},
\\
&& [\, \hat{r}_H\, ]
   \qquad\ \ \rightarrow \qquad
   \hat{r}_{H\alpha}=\kappa\,\hat{r}_{\alpha},
\\
&&  [U^\pm_{_{\!\!HZ}}]
  \qquad\ \ \! \rightarrow \qquad
  U^\pm_{\beta\alpha},
\\
&&  [\,\hat{r}_{H}\,]^n
  \qquad\, \rightarrow \qquad
  \hat{r}_{H\alpha_1}\cdots\hat{r}_{H \alpha_n}
  = \kappa^n \hat{r}_{\alpha_1}\cdots \hat{r}_{\alpha_n},
\\
&&  [U^\pm_{\!H}]^n
  \qquad\,\, \rightarrow \qquad
  U^\pm_{\mu_1\alpha_1} \cdots
  U^\pm_{\mu_n\alpha_n},
\\
&&  [U^\pm_{\!Z}]
  \qquad\,\ \ \rightarrow \qquad
  U^\pm_{\mu\beta}.
\end{eqnarray}
For any operator or re-scaled momentum, its zeroth power term with 
$n=0$ is set to $1$ for convenience.  Without loss of generality, any 
arrangement of the $\alpha$ four-vector indices is ultimately absorbed 
into the symmetric nature of the $H$ wave tensors, rendering all the
permutations of these indices effectively identical. 
Additionally, it's worth noting that the re-scaled momenta, 
$\hat{p}_\star$ and  $\hat{p}_Z$ vanish when their corresponding 
masses, $m_\star$ and $m_Z$, are set to zero such that only the operators corresponding to the maximal-magnitude helicity values survive 
in the massless limit, thereby rendering their theoretical treatment 
automatic and accurate.\s

\setcounter{equation}{0}

\section{Weaving helicity-specific covariant four-point vertices}
\label{sec:weaving_covariant_4-point_vertices}

We are now ready to express the helicity amplitude of the decay
process $H\to \ell^-\ell^+ Z$ in the covariant formulation, 
which can be written in a factorized form as
\begin{eqnarray}
\mathcal{M}^{[H\to\ell\ell Z]}_{\lambda_H;\sigma,
   \lambda_Z}(\theta)
= c^{[H\ell \ell Z]}_{\lambda_H; \sigma, \lambda_Z} 
  \left\{\epsilon^{*\beta}_Z(k_Z,\!\lambda_Z)
   {[{\cal H}^{j_0[HZ]}_{[\lambda_H,\!\lambda_Z]}]}^{\bar{\mu}}_{\bar{\alpha}, \beta}
  \epsilon^{\bar{\alpha}}_H(p_H,\!\lambda_H)\right\}
   \left\{\bar{u}(k_1,\!\mbox{\scriptsize $\frac{\sigma}{2}$})
  {[{\cal H}^{j_0[\ell\ell]}_{[\sigma]}]}_{\bar{\mu}}
  v(k_2,\!-\mbox{\scriptsize $\frac{\sigma}{2}$})\right\},
\label{eq:amplitudes_with_covariant_operators}
\end{eqnarray}
in the $H$ rest frame
with $j_0={\rm max}(|\lambda_H+\lambda_Z|, |\sigma|)$,
where $\bar{\alpha}=\alpha_1\cdots\alpha_{s_H}$ is the collection of 
4-vector indices of the spin-$s_H$ wave tensor of the $H$ boson,
$\beta$ is the four-vector index of the spin-1 $Z$ and
$\bar{\mu}=\mu_1\cdots\mu_{j_0}$. Note that $j_0$ is no less than 
the unity because of $|\sigma|=1$ in the CC case. All the 4-momenta
and helicities are denoted in Fig.$\,$\ref{fig:h_to_llz_rm}.
Contracting the general $\ell\ell$ helicity-specific operator
${\cal H}^{j_0[\ell\ell]}_{[\sigma]}$ with the
general $HZ$ helicity-specific  operator
${\cal H}^{j_0[HZ]}_{[\lambda_H,\lambda_Z]}$
results in a four-point $H\ell\ell Z$ vertex operator 
in an operator form as
\begin{eqnarray}
  [{\cal H}^{j_0[H \ell\ell Z]}_{[\lambda_H; \sigma,\lambda_Z]}]
= c^{[H\ell \ell Z]}_{\lambda_H; \sigma, \lambda_Z} \,\,
  [{\cal H}^{j_0[HZ]}_{[\lambda_H,\lambda_Z]}]
  \cdot
   [{\cal H}^{j_0[\ell\ell]}_{[\sigma]}],
\label{eq:covariant_4-point_vertex}
\end{eqnarray}
with a form factor $c^{[H \ell\ell Z]}_{\lambda_H; \sigma, \lambda_Z}$
for the specific helicity configurations $[\lambda_H;\sigma,\lambda_Z]$.
Each of the form factors depends generally on the invariant products of 
momenta, all of which can be expressed by combining 
$m_\star=\sqrt{k^2}$ and $\hat{l}\cdot \hat{p}_\star = \kappa \cos\theta$.
It implies that any parity-odd combination of the form factors
is proportional to at least the single power of $\kappa\cos\theta$.\s

Let us show how to assemble the general form of the covariant
four-point vertex in Eq.$\,$(\ref{eq:covariant_4-point_vertex}) by use 
of the general $\ell\ell$- and $HZ$-helicity-specific operators, step by step. 
First, each of the general $\ell \ell$ operators must 
be constructed by one basic $\ell\ell$ operator 
$W_\pm$ and a multiple product of the normalized four-vector $\hat{l}$ as
%
\begin{eqnarray}
  \mathcal{H}^{j_0[\ell\ell]\bar{\mu}}_{[\sigma]}
= \omega^{j_0-1}_\star\, W_{\sigma}^{\mu_1}\,\hat{l}^{\mu_2}\cdots
   \hat{l}^{\mu_{j_0}}
  \quad \mbox{with}\quad  
    W^{\mu_1}_\sigma
  =\frac{i}{\sqrt{2} m_H} \gamma^{\mu_1}\,P_\sigma 
   \quad\mbox{and}\quad j_0\geq 1,
\label{eq:general_ee_fermionic_operators}
\end{eqnarray}
with the chiral projection operator $P_\sigma=(1+\sigma\gamma_5)/2$, and the 
abbreviation $\bar{\mu}=\mu_1\mu_2\cdots\mu_{j_0}$. 
For the sake of efficient evaluation and analysis, 
we introduce the contraction between the basic $\ell\ell$ operators 
$W^{\mu_1}_{\sigma}$ and the massless spinors, $u(\ell^-)$ and 
$v(\ell^+)$, so as to read straightforwardly the $\ell\ell$ 
vertex part in the decay amplitude as
\begin{eqnarray}
\bar{u}(k_1,\mbox{$\frac{\sigma}{2}$})
[{\cal H}^{j_0[\ell\ell]}_{[\sigma]}]^{\bar{\mu}}v(k_2,\mbox{$\frac{\sigma}{2}$})
\, =\, \omega^{j_0}_\star \, 
  \hat{l}^{\mu_1}_\sigma\, \hat{l}^{\mu_2}\cdots
  \hat{l}^{\mu_{j_0}},
\end{eqnarray}
where the $j_0$-th power term of $\omega^{j_0}_\star$ is introduced
to properly take into account the kinematical properties of the 
original chiral leptonic vector currents $L_\pm$
in Eq.$\,$(\ref{eq:chiral_ll_vector_current}) and the momentum 
difference $l=k_1-k_2=m_H\omega_\star \hat{l}$ varying with 
$\omega_\star=m_\star/m_H$.\s

\begin{figure}[ht!]
\vskip 0.2cm
\centering
\includegraphics[scale=0.33]{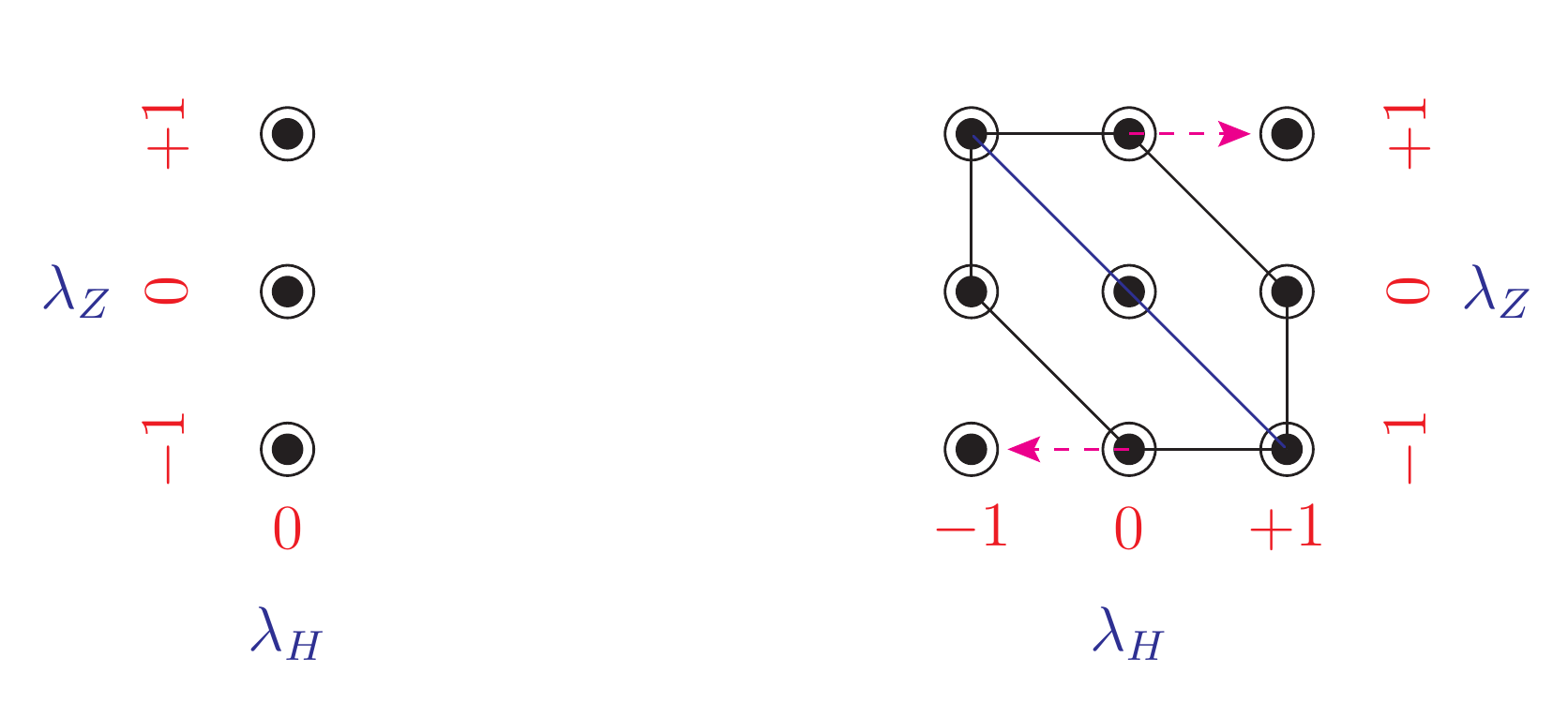}
\vskip -0.2cm
\caption{\it Diagrammatic representation of the lattice structure of 
   the helicity space for both the $H$ and $Z$ particles in scenarios
   involving a spin-0 $H$ boson (on the left) and a spin-1 $H$ boson 
   (on the right). For illustration, three (nine) points
   are displayed for the spin-0 (spin-1) case. In the right diagram,
   the seven points connected through solid lines correspond to the
   total spin $j_0=1$ mode, while the two points connected by the magenta 
   dashed arrows correspond to the total spin $j_0=2$ mode. 
   The value of $j_0$ is one plus the total number of dashed lines 
   required to reach the point from each reference point.
 }
\label{fig:HZ_helicity_lattice_space_spin-0_spin_1}
\end{figure}

Certainly, the simplest case for the decay process $H\to\ell^-\ell^+ Z$ 
is when $s_H=0$, i.e. $H$ is a spin-0 scalar like the SM Higgs boson 
with no helicity at all. The $Z$-boson spin is $s_Z=1$
so that the value of $j_0={\rm max}(|\lambda_Z|,|\sigma|)$ is 
also 1. In this case, the $\ell\ell$ operators 
are $[{\cal H}^{1[\ell\ell]}_{[\sigma]}]^\mu 
=W_\sigma^\mu$ and
the three independent $HZ$-helicity-specific operators are
\begin{eqnarray}
{[{\cal H}^{1[HZ]}_{[\bullet,\pm 1]}]}_{\beta\mu}
   = U^\pm_{\beta\mu} =U^\mp_{\mu\beta}
   \qquad \mbox{and}\qquad
{[{\cal H}^{1[HZ]}_{[\bullet, 0]}]}_{\beta\mu} 
   = U^0_{\beta\mu}
   = \hat{p}_{\star\mu} \hat{p}_{Z\beta}
  =  4\omega_\star \omega_Z \, \hat{p}_\mu \hat{p}_\beta,
\label{eq:spin-0_HZ_operators}
\end{eqnarray}
of which the corresponding three helicity points are displaced in the 
left panel of 
Fig.$\,$\ref{fig:HZ_helicity_lattice_space_spin-0_spin_1}.
The bullet notation $\bullet$ in Eq.$\,$\eqref{eq:spin-0_HZ_operators} 
is used to indicate the spinless nature
of the $H$ boson with $s_H=0$.
Consequently, combining the operators, $[{\cal H}^{1[\ell\ell]}_{[\sigma]}]$ and $[{\cal H}^{1[HZ]}_{[\bullet,\lambda_Z]}]$, 
over the $\mu$ index, we have six ($6=2\times 3$) 
independent terms with the 
Lorentz-invariant $m_\star$- and $\hat{l}\cdot\hat{p}_\star$-dependent 
coefficients. Note that $\hat{l}\cdot\hat{p}_\star = \kappa\cos\theta$ 
in the $e^-e^+$ CM frame in Fig.$\,$\ref{fig:h_to_llz_rm}.\s

When the spin of the massive $H$ boson is $s_H=1$, there are nine
$HZ$ operators for each of two $\ell\ell$ operators, as described in 
the right frame of
Fig.$\,$\ref{fig:HZ_helicity_lattice_space_spin-0_spin_1} 
and listed in Tab.$\,$\ref{tab:spin-1_covariant_operators}. 
Consequently, there are
in total eighteen independent terms with proper Lorentz-invariant
$m_\star$- and $\hat{l}\cdot\hat{p}_\star$-dependent coefficients.
The invariant product $\hat{l}\cdot\hat{p}_\star=\kappa \cos\theta$
is always accompanied by a kinematical factor $\kappa$. \s

\begin{table}[htb!]
\vskip 0.5cm
\centering
\begin{tabular}{||c|c|c|c|c|c||}
\hline
   &  &  &  &  &
\\[-0.4cm]
$s_H$ & $j_0$ & $(\lambda_H, \lambda_Z)$
      & $[{\cal H}^{j_0[\ell\ell]}_{[\sigma]}]$
      & $[{\cal H}^{j_0[HZ]}_{[\lambda_H,\lambda_Z]}]$
      & $\#$
\\[-0.4cm]
  &  &  &  &  &
\\ \hline
      & \rb{-2.3ex}{$1$}
      & \rb{-0.8ex}{$(\pm 1, \mp 1), (0,0)$}
      & \rb{-2.3ex}{$W^{\mu_1}_\sigma $}
      & \rb{-0,8ex}{$U^\pm_{\beta\alpha}\hat{p}_{\star\mu_1},\ \
                \hat{r}_{H\alpha}\hat{p}_{\star\mu_1} \hat{p}_{Z\beta}$}
      & \rb{-0.8ex}{$3$}
\\
      &
      & $(\pm 1, 0), (0, \pm 1)$
      &
      & $U^\pm_{\mu_1\alpha}\hat{p}_{Z\beta},
          \ \
          U^\pm_{\mu_1\beta}\hat{r}_{H\alpha}$
      & $4$
\\[0.1cm] \cline{2-6}
\rb{2.5ex}{$1$}  & \rb{-1.0ex}{$2$}
      & \rb{-1.0ex}{$(\pm 1, \pm 1)$}
      & \rb{-1.0ex}{$\omega_\star W^{\mu_1}_\sigma \hat{l}^{\mu_2}$}
      & \rb{-1.0ex}{$U^\pm_{\mu_1\alpha}U^\pm_{\mu_2\beta}$}
      & \rb{-1.0ex}{$2$}
\\[0.4cm] \hline
\end{tabular}
\mbox{ }\\[0.3cm]
\caption{\it The table of two $\ell\ell$-helicity-specific operators 
   with $\sigma = \pm 1$ and nine $HZ$-helicity-specific operators 
   with a massive spin-1 particle $H$ classified according to the
   values of $j_0$ and 
   each  $(\lambda_H,\lambda_Z)$ helicity combination for the spin-1 $H$.
}
\label{tab:spin-1_covariant_operators}
\end{table}

\begin{figure}[ht!]
\vskip 0.1cm
\centering
\includegraphics[scale=1.20]{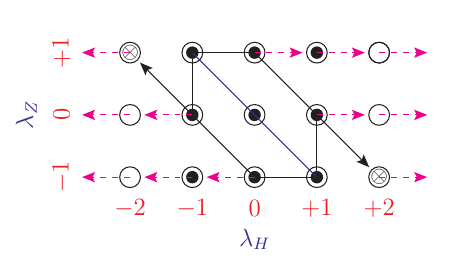}
\caption{\it Diagrammatic description of the lattice structure of
   the $H$ and $Z$ helicity space. For illustration, fifteen points
   are displayed for the case up to $s_H=2$. The nine points connected
   through solid lines correspond to the $j_0=1$ mode,
   while the points connected by magenta dashed arrows correspond to the $j_0>1$ 
   modes. 
   The value of $j_0$ is one plus the total number
   of dashed arrows required to reach the point from each starting 
   reference point. For example, the $j_0$ value of the point
   $(\lambda_H,\lambda_Z)=(\pm 2, \pm 1)$ is $3=|\pm 2 \pm 1|$ as 
   it involves two dashed lines. 
   }
\label{fig:spin-2_HZ_helicity_lattice_space}
\end{figure}

When the spin of the $H$ boson is $s_H=2$, there are fifteen $HZ$-helicity-specific operators for each of two $\ell\ell$-helicity-specific operators, as listed 
in Tab.$\,$\ref{tab:spin-2_covariant_operators}. 
Its pictorial description can be inferred from the diagram in 
Fig.$\,$\ref{fig:spin-2_HZ_helicity_lattice_space}, 
restricting $|\lambda_H|\leq 2$.
Consequently, there are in total thirty independent terms with proper
Lorentz-invariant $m_\star$- and $\hat{l}\cdot\hat{p}_\star$-dependent
coefficients.\s

\begin{table}[htb!]
\vskip 0.5cm
\centering
\begin{tabular}{||c|c|c|c|c|c||}
\hline
   &  &  &  &  &
\\[-0.4cm]
$s_H$ & $j_0$ & $(\lambda_H, \lambda_Z)$
      & $[{\cal H}^{j_0[\ell\ell]}_{[\sigma]}]$
      & $[{\cal H}^{j_0[HZ]}_{[\lambda_H,\lambda_Z]}]$
      & $\#$
\\[-0.4cm]
  &  &  &  &  &
\\ \hline
      & \rb{-2.3ex}{$1$}
      & \rb{-0.8ex}{$(\pm 1, \mp 1), (0,0)$}
      & \rb{-2.3ex}{$ W^{\mu_1}_\sigma $}
      & \rb{-0,8ex}{$ U^\pm_{\alpha_1\beta} \hat{r}_{H\alpha_2}
                      \hat{p}_{\star \mu_1},\ \
                     \hat{r}_{H\alpha_1} \hat{r}_{H\alpha_2}
                     \hat{p}_{\star \mu_1}\hat{p}_{Z\beta}$}
      & \rb{-0.8ex}{$3$}
\\
      &
      & $(\pm 2, \mp 1), (\pm 1, 0), (0, \pm 1)$
      &
      & $ U^\pm_{\mu_1\alpha_1}U^\pm_{\beta\alpha_2},\ \
          U^\pm_{\mu_1\alpha_1}\hat{r}_{H\alpha_2} \hat{p}_{Z\beta},\ \
          U^\pm_{\mu_1\beta}\hat{r}_{H\alpha_1}\hat{r}_{H\alpha_2}$
      & $6$
\\[0.1cm] \cline{2-6}
\rb{-2.5ex}{$2$}
      & \rb{-2.3ex}{$2$}
      & \rb{-1.0ex}{$(\pm 1, \pm 1)$}
      & \rb{-2.3ex}{$\omega_\star W^{\mu_1}_\sigma \hat{l}^{\mu_2}$}
      & \rb{-1.0ex}{$U^\pm_{\mu_1\alpha_1}U^\pm_{\mu_2\beta}
                    \hat{r}_{H\alpha_2}$}
      & \rb{-1.0ex}{$2$}
\\
      &
      & $(\pm 2, 0)$
      &
      & $U^\pm_{\mu_1\alpha_1}U^\pm_{\mu_2\alpha_2} \hat{p}_{Z\beta}$
      & $2$
\\[0.1cm]\cline{2-6}
      & \rb{-1.0ex}{$3$}
      & \rb{-1.0ex}{$(\pm 2, \pm 1)$}
      & \rb{-1.0ex}{$\omega^2_\star W^{\mu_1}_\sigma
                     \hat{l}^{\mu_2}\hat{l}^{\mu_3}$}
      & \rb{-1.0ex}{$U^\pm_{\mu_1\alpha_1} U^\pm_{\mu_2\alpha_2}
                     U^\pm_{\mu_3\beta}$}
      & \rb{-1.0ex}{$2$}
\\[0.4cm] \hline
\end{tabular}
\mbox{ }\\[0.3cm]
\caption{\it The table of two $\ell\ell$-helicity-specific 
    operators with $\sigma=\pm 1$ and fifteen 
    $HZ$-helicity-specific operators with a spin-2 massive particle $H$,
    classified according to three allowed values of $j_0=1,2,3$
    and each available $(\lambda_H,\lambda_Z)$ helicity combination.
}
\label{tab:spin-2_covariant_operators}
\end{table}

Finally,  if the $H$ spin is $s_H\geq 3$, there are  in total $3(2s_H+1)$ 
$HZ$ operators for each of two $\ell\ell$ operators as shown in Fig.$\,$\ref{fig:spin-2_HZ_helicity_lattice_space}.  
The $j_0=1$ category has nine independent $HZ$ terms. In addition,
for each $Z$ helicity, $\lambda_Z=\pm 1, 0$  there are $2(s_H-1)$ 
independent $HZ$ for all the $j_0>1$ categories. Summing them, 
we have $3\times 2(s_H-1) + 9 = 3(2s_H+1)$ as expected. 
Consequently, there are in total $6(2s_H+1)$ independent terms 
with Lorentz-invariant coefficients dependent only on
$m_\star$ and $\hat{l}\cdot\hat{p}_\star$. \s

For a given $j_0$, the general CC $\ell\ell$ operators are nothing but 
$[{\cal H}^{j_0[\ell\ell]}_{[\sigma]}]$ of which the expression was 
given previously in Eq.$\,$(\ref{eq:general_ee_fermionic_operators}). 
They can be cast into the following operator form
\begin{eqnarray}
  [{\cal H}^{j_0[\ell\ell]}_{[\sigma]}]
= 
\omega^{j_0-1}_\star\, [W_{\sigma}][\,\hat{l}\,]^{j_0-1}\quad
  \Rightarrow\quad
  {\cal H}^{j_0[\ell\ell]\bar{\mu}}_{[\sigma]}
= \omega^{j_0-1}_\star\, W^{\mu_1}_{\sigma}\hat{l}^{\mu_2}\cdots \hat{l}^{\mu_{j_0}},
\end{eqnarray}
with $\bar{\mu}=\mu_1\cdots\mu_{j_0}$. We emphasize again that 
the $\omega^{j_0-1}_\star$ term is introduced to take into account the
dependence of the momentum difference $l=k_1-k_2$ on the varying 
$\ell^-\ell^+$ invariant mass $m_\star$.\s


Note that all the $HZ$-helicity-specific operators related to the two-body decay of the $H$ boson include only two independent normalized momenta, $\hat{p}$ and $\hat{r}$. 
The $HZ$-helicity-specific operators are categorized into six parts as 
\begin{eqnarray}
 \big[\mathcal{H}^{j_0[HZ]}_{[0,0]}\big]
 &=& 
    [\,\hat{r}_{H}\,]^{s_H} [\;\hat{p}_\star\;] 
    [\,\hat{p}_{Z}\,] 
  \qquad\quad\quad  \mbox{with}\quad  j_0=1,
\label{eq:hz_00_operator}
\\[1mm]
 \big[\mathcal{H}^{j_0[HZ]}_{[0, \pm 1]}\big]
 &=&  [\,\hat{r}_{H}\,]^{s_H}[U^{\pm}_{\!Z}] 
  \qquad\quad\quad\qquad\, \mbox{with}\quad j_0=1,
\label{eq:hz_11_operator}
\\[1mm]
 \big[\mathcal{H}^{j_0[HZ]}_{[\pm 1 , \mp 1]}\big]
 &=&   [\,\hat{r}_{H}\,]^{s_H-1}\,
    [U^{\pm}_{_{\!\!HZ}}] [\;\hat{p}_\star\;]
 \quad\quad\quad \mbox{with }\quad  j_0=1,
\label{eq:hz_01_operator}
\\[1mm]
 \big[\mathcal{H}^{j_0[HZ]}_{[\pm n, 0]}\big]
 &=&   [\,\hat{r}_{H}\,]^{s_H-n}\,
    [U^{\pm}_{\!H}]^{n} [\,\hat{p}_{Z}\,]
  \quad\quad\quad\!\! \mbox{with }\quad j_0=n,
\label{eq:hz_n0_operator}
\\[1mm]
 \big[\mathcal{H}^{j_0[HZ]}_{[\pm \tilde{n}, \mp 1]}\big]
 &=&  [\,\hat{r}_{H}\,]^{s_H-\tilde{n}}\,
    [U^{\pm}_{\!H}]^{\tilde{n}-1} 
    [U^\pm_{_{\!\!HZ}}] 
  \quad\quad\!\! \mbox{with }\quad j_0=\tilde{n}-1,
\label{eq:hz_n_tilde_1_operator}
\\[1mm]
 \big[\mathcal{H}^{j_0[HZ]}_{[\pm n, \pm 1]}\big]
  &=& [\,\hat{r}_{H}\,]^{s_H-n}\,
   [U^{\pm}_{\!H}]^{n}  [U^{\pm}_{\!Z}]
  \qquad\quad\! \mbox{with }\quad j_0=n+1,
\label{eq:hz_n1_operator}
\end{eqnarray}
where $n=1,\cdots, s_H$ and $\tilde{n}=2,\cdots, s_H$. In total
we can compute the total number of independent $HZ$ operators 
by summing the degrees of freedom of all the categories
as
\begin{eqnarray}
n[s_H] = 1+2+2+2 s_H + 2 (s_H-1) + 2s_H = 3\, (2s_H+1),
\end{eqnarray}
as expected. The first factor of 3 in the last expression reflects 
that the $Z$ boson is a spin-1 massive particle with $3=2\times 1+1$
degrees of freedom.\s

Equivalently, the CC helicity amplitude
\eqref{eq:amplitudes_with_covariant_operators}
of the decay process $H\to \ell^-\ell^+ Z$ in the rest frame of 
the $H$ boson can be generally decomposed in the helicity 
formalism~\cite{Wick:1962zz,Dreiner:2008tw,
Dreiner:2023yus,Martin:1970hmp,Leader:2011vwq, Choi:2019aig} into one
dynamical part $\mathcal{T}$ and a pure 
angular part denoted by a Wigner $d$ function as
\begin{eqnarray}
\mathcal{M}^{[H\to \ell\ell Z]}_{\lambda_H; \sigma, \lambda_Z}(\theta)
= \mathcal{T}^{[H\ell\ell Z]}_{\lambda_H; \sigma, \lambda_Z}
   (m_\star, \kappa\cos\theta)\;
  d^{j_0}_{\lambda_H+\lambda_Z, \sigma}(\theta),
\label{eq:3-body_cc_decay_helicity_amplitude}
\end{eqnarray}
in terms of the polar angle $\theta$ defining the $\ell^-$ momentum direction
of the $\ell^-\ell^+$ CM system with respect to the positive $z$-axis,
where $j_0= \mbox{max}(|\lambda_H+\lambda_Z|, |\sigma|)$ is the minimal 
value of the angular momenta of the $\ell^-\ell^+$ CM system for the
fixed helicity configuration.\footnote{
$d^j_{\lambda_H+\lambda_Z, \sigma}(\theta)$ with 
$j> j_0$ is a simple product of $d^{j_0}_{\lambda_H+\lambda_Z, \sigma}(\theta)$ and a $(j-j_0)$-th order polynomial of $\cos\theta$.}
The explicit form of the Wigner $d$ function can be found in several
standard angular-momentum textbooks, for instance, in 
Ref.~\cite{Rose:1957aa}. 
For convenience, a comprehensive set of analytic formulas useful 
for systematically and efficiently calculating the helicity amplitudes 
is provided in Appendix~\ref{appendix:analytic_formulas}.\s

It is worthwhile to emphasize that the reduced helicity amplitudes  
$\mathcal{T}^{[H\ell\ell Z]}_{\lambda_H; \sigma, \lambda_Z}$ depend 
at most on the $\ell^-\ell^+$ invariant mass $m_\star$ and the cosine 
of the polar-angle, $\cos\theta$, accompanied without exception by $\kappa$ 
in the combination of $\kappa\cos\theta$.
If the form factors are for any contact vertices, they are independent of 
the angular variable. In general, their behaviors rely on the three 
Lorentz-invariant Mandelstam variables~\cite{Mandelstam:1958xc},
which are defined to be
\begin{eqnarray}
&& s = (p_H-k_Z)^2= (k_1+k_2)^2 = m^2_\star,\\ 
&& t = (p_H-k_1)^2 \, = (k_Z+k_2)^2
     = (m^2_H + m^2_Z-m^2_\star- m^2_H \kappa\cos\theta)/2,\\
&& u = (p_H-k_2)^2 = (k_Z+k_1)^2
     = (m^2_H + m^2_Z-m^2_\star + m^2_H \kappa\cos\theta)/2,
\end{eqnarray}
with their explicit form derived in the $H$ rest frame. As argued 
previously, the Mandelstam variables take 
$s=(m_H-m_Z)^2$ and $t= u = m_Hm_Z$ at the threshold with 
$m_\star=m_H-m_Z$, being independent of 
the polar angle $\theta$ indicating that the $Z$ boson and the 
$\ell^-\ell^+$ system are produced at rest.\s

The specific form  of the Wigner $d$ function in 
Eq.$\,$(\ref{eq:3-body_cc_decay_helicity_amplitude}) is guaranteed by 
the characteristic feature that any helicity is Lorentz invariant
with trivial Wick helicity rotation in the massless limit 
so that the decay helicity amplitude in the $H$ rest frame is 
identical to that in the $\ell^-\ell^+$ CM 
frame~\cite{Martin:1970hmp,Leader:2011vwq,Choi:2019aig}. 
The Wigner $d$ function takes the following explicit 
forms~\cite{Rose:1957aa}:
\begin{eqnarray}
  d^{j_0}_{\pm j_0,\sigma}(\theta)
= (-1)^{j_0-1} \, \sqrt{\frac{(2j_0)!}{(j_0+1)!(j_0-1)!}}\,
   \left(\frac{\sin\theta}{2}\right)^{j_0-1}\, d^1_{\pm 1, \sigma}(\theta),
\label{eq:l-dependent_d_functions} 
\end{eqnarray}
for $j_0=|\lambda_H+\lambda_Z|\geq 1$ with 
$d^1_{\pm 1, \sigma}(\theta) = (1+\sigma\cos\theta)/2$ and 
it is necessary to introduce one additional Wigner $d$ function 
\begin{eqnarray}
d^1_{0,\sigma}(\theta) = \frac{\sigma}{\sqrt{2}}\sin\theta,
\end{eqnarray}
for $j_0=|\sigma|=1$ when $\lambda_H+\lambda_Z=0$.\s

As mentioned before, the $H$ polarization states will be  
averaged out eventually in evaluating every physical observable 
under consideration so that the $H$ polarization axis
can be set to be opposite to the flight direction of the $Z$ boson, 
as described in Fig.~\ref{fig:h_to_llz_rm}, for much more 
efficient analytic calculation. Nevertheless, 
if required, the $H$ decay process can be connected straightforwardly 
with its production mechanism by explicitly specifying the $H$ polarization
axis with its own polar and azimuthal angles, $\theta_H$ and $\phi_H$
as well as an azimuthal angle $\phi$ for denoting the coordinate-independent
geometrical orientation of the $\ell^-\ell^+$ system
with respect to the polarization axis of the $H$ boson.  
In this case, the three-body decay helicity amplitudes of the $H$ boson 
is written as
\begin{eqnarray}
  {\cal M}^{[H\to\ell\ell Z]}_{\lambda_H;\sigma,\lambda_Z}
  (\theta_H, \phi_H; \theta,\phi)
= \sum^{s_H}_{\lambda'_H=-s_H}
   D^{s_H*}_{\lambda'_H,\lambda_H}(\pi+\phi,\theta_H,\phi_H)\,
  \mathcal{M}^{[H\to \ell\ell Z]}_{\lambda'_H; \sigma, \lambda_Z}(\theta),
\end{eqnarray}
with the simplified
decay helicity amplitude 
$\mathcal{M}^{[H\to \ell\ell Z]}_{\lambda'_H; \sigma, \lambda_Z}$ 
in Eq.~\eqref{eq:3-body_cc_decay_helicity_amplitude}, where 
the rotation matrix  elements $D^{s_H*}_{\lambda'_H,\lambda_H}$ are 
defined to be
\begin{eqnarray}
  D^{s_H*}_{\lambda'_H,\lambda_H}(\pi+\phi,\theta_H,\phi_H)
  =e^{i\lambda_H'(\pi+\phi)}\,d^{s_H}_{\lambda_H',\lambda_H}(\theta_H)\,
   e^{i\lambda_H \phi_H},
\end{eqnarray}
which satisfies the orthogonality property for the solid-angle integration
on the unit sphere as
\begin{eqnarray}
\int d\phi \,d\Omega_H 
    D^{s_H*}_{\lambda'_H, \lambda_H}(\pi+\phi,\theta_H,\phi_H)\,
    D^{s_H}_{\lambda''_H.\lambda_H}(\pi+\phi,\theta_H,\phi_H)
= \frac{8\pi^2}{2s_H+1}\,\delta_{\lambda'_H,\lambda''_H},
\end{eqnarray}
with the differential solid-angle measure
$d\Omega_H=\sin\theta_H d\theta_H d\phi_H$.\s

We emphasize that the reduced helicity amplitudes 
$\mathcal{T}^{[H\ell\ell Z]}_{\lambda_H;\sigma,\lambda_Z}(m_\star, 
\kappa\cos\theta)$ do not require the
introduction of any cut-off scale $\Lambda$, probably 
indicating the energy scale of a new physics. Our theoretical description
operates beyond the limitations of low-energy regimes by allowing 
the form factors $c^{[H\ell\ell Z]}$ in Eq.~\eqref{eq:amplitudes_with_covariant_operators} to be functions 
of particle momenta without restriction to low powers. 
One notable common feature is the polar-angle $\theta$ independence of 
all the form factors at the threshold 
with $m_\star = m_H-m_Z$ rendering the value of $\kappa$ zero. 
In this regard, the pure angular component denoted by the Wigner $d$ 
function may be regarded as
the minimal angular distribution for any given helicity combination.\s

\setcounter{equation}{0}

\section{Characteristic decay patterns of the SM Higgs boson}
\label{sec:sm_higgs_decays}

As mentioned before, the spin $s_H$, parity $P$ and charge conjugation 
$C$ quantum numbers of the SM Higgs boson are necessarily $s_H[PC]=0[++]$  
by the construction of the SM with the EW gauge symmetry hidden or 
spontaneously broken~\cite{Weinberg:1967tq,Salam:1968rm}. 
Therefore, establishing the basic structure of the SM requires
identifying the spin and parity of the Higgs state  
as well as other characteristic features through as many available 
processes as possible  at the ongoing LHC and future high-energy lepton 
and hadron collider experiments.\s

One of the prime Higgs decay modes for its spin and parity identification, 
which actually has played a crucial role in discovering the Higgs 
boson~\cite{ATLAS:2012yve,CMS:2012qbp},
is the 3-body decay process of the SM Higgs boson $H\to \ell^-\ell^+ Z$ 
proceeding through the production $H\to Z^\star Z$ of a virtual boson 
$Z^\star$ and its sequential two-body leptonic decay $Z^\star\to\ell^-\ell^+$
with $\ell=e, \mu$:
\begin{eqnarray}
H\ \ \to\ \ Z^\star + Z \ \ \to \ \ \ell^-\ell^+ + Z,
\end{eqnarray}
where the virtual $Z^\star$ can be treated as an on-shell spin-1 particle
of varying mass $m_\star=\sqrt{k^2}$ in the CC limit of neglecting the lepton
masses. In the kinematic configuration in Fig.$\,$\ref{fig:h_to_llz_rm},
the polarization vectors of the off-shell $Z^\star$ boson,
whose the polarization axis is set to be along the positive $z$-axis,
are given by
\begin{eqnarray}
 \epsilon_\star(k,\pm 1)
=\frac{1}{\sqrt{2}}(0, \mp 1, -i,0),\qquad
 \epsilon_*(k, 0)
= \frac{1}{2\omega_\star}(\kappa, 0, 0, e_\star),
\end{eqnarray}
so that the helicity amplitudes of the three-body decay 
$H\to\ell^-\ell^+ Z$ can be decomposed into one $Z^\star$ 
production part and one $Z^\star$ decay part connected by 
a virtual $Z^\star$ propagator as
\begin{eqnarray}
   {\cal M}[H\to Z^\star Z\to \ell^-\ell^+] 
 =  \frac{1}{m^2_\star-m^2_Z+i m_Z\Gamma_Z}\, 
   {\cal P}[H\to Z^\star Z]\,
   {\cal D}[Z^\star \to \ell^-\ell^+],
\end{eqnarray}
where the helicity amplitudes for the two-body decay $H\to Z^\star Z$ 
and the other two-body decay $Z^\star \to\ell^-\ell^+$ explicitly read
\begin{eqnarray}
{\cal P}[H\to Z^\star ({\lambda_\star}) Z(\lambda_Z)]
 \! &=&\! g_Z\, m_Z\, 
    [\epsilon^*(k,\lambda_\star)\cdot \epsilon^*(k_Z, \lambda_Z)]
     \, =\, g_Z\, m_Z\, P_{\lambda_Z, \lambda_Z},
           \delta_{\lambda_\star, \lambda_Z}\\
{\cal D}\big[Z^\star ({\lambda_\star})\to 
        \ell^-(\mbox{$\frac{\sigma}{2}$})
        \ell^+(\mbox{$-\frac{\sigma}{2}$})\big]
 \! &=&\!\!   i\sqrt{2} g_Z\, m_\star\,
         [v_\sigma \hat{l}_\sigma\cdot \epsilon(k, \lambda_\star)]
     \quad \ \ 
     \, =\, -i\sqrt{2} g_Z\, m_\star\, v^\ell_\sigma\,
        D_{\lambda_\star, \sigma} ,
\end{eqnarray}
with $g_Z = e/c_W s_W$,  $v_+=s^2_W$ and $v_-=s^2_W-1/2$ 
in terms of the proton electric charge $e$ and the cosine and 
sine,  $c_W=\cos\theta_W$ and $s_W=\sin\theta_W$,  
of the weak mixing angle $\theta_W$. The helicity values $\lambda_\star$ and 
$\lambda_Z$ of the off-shell $Z^\star$ and on-shell $Z$ bosons are $\pm 1$ 
or $0$. Note that the imaginary number $i$ is explicitly
taken out to set the normalized chiral lepton current 
$\hat{l}_\sigma$ to be a pure spin-1 polarization vector in 
the Wick convention. \s

The metric tensor $g_{\beta\mu}$ of even normality 
$n_H\equiv (-1)^{s_H}\, P=+$ defines the pure scalar coupling of 
the SM Higgs boson to two $Z^{(*)}$ bosons. This coupling is generated
characteristically through EWSB and can be effectively decomposed 
in terms of fundamental $HZ$ operators and re-scaled momenta as
\begin{eqnarray}
  \langle Z^\star_\mu Z_\beta|H\rangle_{\rm SM} 
= g_Z m_Z g_{\mu\beta}
  \ \ \Rightarrow\ \ 
 g_{\mu\beta}
= U^+_{\mu\beta}+U^-_{\mu\beta}
 + \frac{1-\omega^2_\star-\omega^2_Z}{2\omega_\star\omega_Z\,\kappa^2} 
  \, \hat{p}_{\star \mu}\, \hat{p}_{Z\beta}\,.
\label{eq:metric_decomposition}
\end{eqnarray}
The decomposition (\ref{eq:metric_decomposition}) enables us to derive 
the helicity amplitudes $P_{\lambda_\star, \lambda_Z}$ straightforwardly, 
of which the three non-zero $Z^\star$ production amplitudes read
\begin{eqnarray}
  P_{\pm 1, \pm 1}
= -1,\qquad
  P_{0,0}
= \frac{1-\omega^2_Z-\omega^2_\star}{2\omega_Z \omega_\star}
=  \frac{m^2_H-m^2_Z-m^2_\star}{2m_Zm_\star}.
\end{eqnarray}
On the other hand, the helicity amplitudes of the two-body decay 
$Z^\star \to \ell^-\ell^+$ can be cast into the pure kinematical 
angular-dependent form:  
\begin{eqnarray}
  D_{\pm 1,\sigma}
= \frac{1\pm \sigma\cos\theta}{2}
= d^1_{\pm 1, \sigma}(\theta)
 ,\quad
  D_{0, \sigma}
= \frac{\sigma}{\sqrt{2}}\sin\theta
= d^1_{0,\sigma}(\theta),
\end{eqnarray}
apart from the chiral vector coupling $v_\sigma$ with $\sigma=\pm 1$, 
in terms of the polar angle $\theta$ denoting
the $\ell^-$ flight direction in the $\ell^-\ell^+$ CM frame.
Note that the $Z^* \rightarrow \ell^-\ell^+$ decay amplitudes are 
invariant under the boost along the $Z^*$ polarization axis, 
as guaranteed from the feature that the helicity of
a massless particle is a Lorentz-invariant quantity.\s

Based on the explicit forms of the decay helicity amplitudes, we extract
and summarize the essential points for measuring the spin and parity
of the SM Higgs boson. Firstly,  the partial decay width of the process
$H\to Z^\star Z$ is given in the SM by
\begin{eqnarray}
  \Gamma(H\to Z^\star Z)
= \frac{3G^2_F m^4_Z}{8\pi^3} \delta_Z m_H R(\omega_Z),
\end{eqnarray}
where $\delta_Z = 7/12-10 s^2_W/9+40 s^4_W/27$ and the $\omega_Z$-dependent function
$R(\omega_Z)$ is decomposed as
\begin{eqnarray}
   R(\omega_Z)
= (1-4\omega_Z^2+12\omega^4_Z) A_0(\omega_Z)
        -2(1-6\omega_Z^2)A_1(\omega_Z)
        +A_2(\omega_Z),
\end{eqnarray}
of which the expression is given in the limit of neglecting the $Z$-boson width 
$\Gamma_Z$ by~\cite{Bredenstein:2006nk,
Keung:1984hn,Barger:1993wt,Djouadi:1997yw,Djouadi:2018xqq}
\begin{eqnarray}
  R(\omega_Z)
&=& \frac{3(1-8\omega_Z^2+20\omega_Z^4)}{2\sqrt{4\omega_Z^2-1}}\cos^{-1}
  \left(\frac{3\omega_Z^2-1}{2\omega_Z^3}\right)
   -\frac{1-\omega_Z^2}{4\omega_Z^2}(2-13\omega_Z^2+47\omega_Z^4)
   \nonumber\\
&& - \frac32(1-6\omega_Z^2+4\omega_Z^4)\log \omega_Z.
 \label{eq:function_r_x}
\end{eqnarray}
The integral-form definition of the $\omega_Z$-dependent
functions $A_{n}(\omega_Z)$ and their explicit forms from $n=0$ to $5$
in the limit of neglecting the $Z$-boson width
can be found in Appendix~\ref{appedix:integral_functions}.
The invariant-mass spectrum $m_\star$ of the off-shell $Z^\star$ boson 
exhibits a rapid rise as $m_\star$ increases, followed by a sharp decline
near the kinematic limit where $m_\star = m_H - m_Z$, signifying zero momentum of the off-shell $Z^\star$ and on-shell $Z$ bosons 
in the final state. Its precise expression is given by
\begin{eqnarray}
  \frac{d\Gamma_H}{dm_\star}
= \frac{3G^2_F m^4_Z\delta_Z}{8\pi^3 m_H} \cdot
  \frac{m_\star(12m^2_\star m^2_Z+m^4_H \kappa^2)}{
       (m^2_\star-m^2_Z)^2+m^2_Z\Gamma^2_Z}\,
  \kappa = \frac{\Gamma_H}{m_H R(\omega_Z)}\,
   \frac{\omega_\star(12\omega^2_\star \omega^2_Z+\kappa^2)}{
       (\omega^2_\star-\omega^2_Z)^2+\omega^2_Z\Gamma^2_Z/m^2_H}\kappa ,
\label{eq:sm_higgs_invariant_mass_distribution}
\end{eqnarray}
where $\kappa$ is the magnitude of the $Z$ and $Z^\star$ momenta in units
of $m_H/2$ in the $H$ rest frame, i.e. $\kappa^2=[1-(\omega_Z+\omega_\star)^2]
[(1-(\omega_Z-\omega_\star)^2]$. The invariant-mass spectrum decreases linearly
with $\kappa$ and therefore steeply with the invariant mass $m_\star$ just
below the threshold:
\begin{eqnarray}
\frac{d\Gamma_H}{dm_\star}\, \sim\, \kappa\, \sim\,
\sqrt{m_H-m_Z-m_\star}\,\,.
\end{eqnarray}
This steep decrease near the threshold is characteristic of the decay of 
the SM Higgs boson into two massive vector bosons of which the 
dimensionless coupling is generated solely through EWSB. \s

\begin{figure}[ht!]
\vskip 0.5cm
\centering
\includegraphics[scale=0.53]{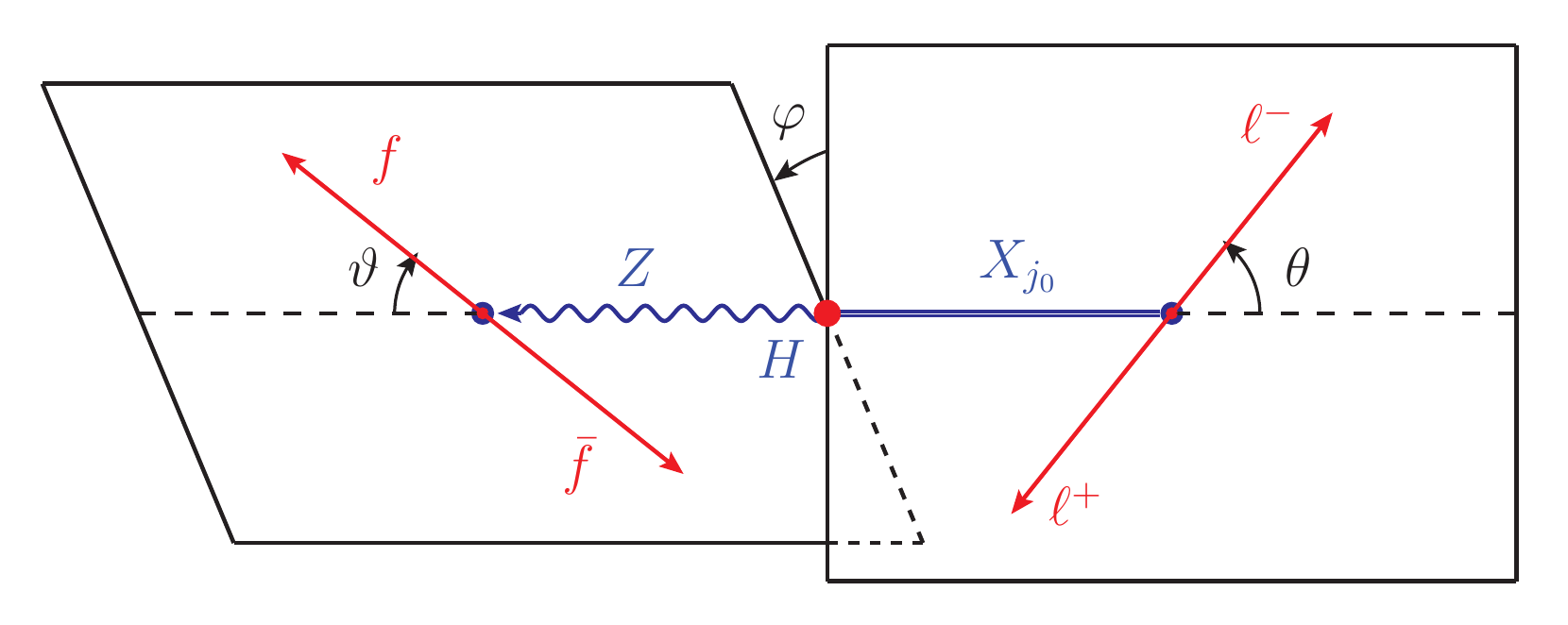}
\vskip 0.3cm
\caption{\it The definition of the polar angles, $\theta$ and $\vartheta$,
   and the azimuthal angle $\varphi$ for the sequential decay
   $H \rightarrow \ell^-\ell^+ Z \rightarrow (\ell^- \ell^+)
   (f \bar{f})$ with $\ell=e, \mu$ and $f=e^-, \mu^-$ in the rest frame of the $H$
   particle. The polar angle
   $\theta$ is defined in the rest frame of the $\ell^-\ell^+$ system
   and the polar angle $\vartheta$ of the lepton $f$ with respect to the
   $Z$-boson flight direction in the rest frame
   of the on-shell $Z$ boson. The azimuthal angle $\varphi$ is the relative angle between two decay planes. The notation  $X$ is introduced for 
   denoting the mixed quantum state of the $\ell^-\ell^+$ system collectively. 
   The value of the allowed angular momentum $j_0$ is fixed to 1 for the 
   intermediate virtual $Z^\star$ in the CC limit of massless leptons, 
   but it generally can take a positive integer from 1 up to $s_H+1$.
   }
\label{fig:angles}
\end{figure}

Correlated polar and azimuthal angle distributions give independent and complementary
access to spin and parity of the $H$ boson as well. Denoting the two 
polar angles of the lepton $\ell$ and the 
fermion $f$ in the decays $Z^\star\to\ell^-\ell^+$ 
and $Z\to f\bar{f}$ in the rest frame of the virtual $Z^\star$ and 
on-shell $Z$ states by $\theta$ and $\vartheta$, respectively, and 
the azimuthal  angle between the lepton-pair plane and the 
fermion-pair plane by $\varphi$ as depicted in Fig.$\,$\ref{fig:angles}, 
the correlated distribution in $\cos\theta$ and $\cos\vartheta$ of 
the off/on-shell $Z^\star$ and $Z$ bosons in the SM is predicted to be
\begin{eqnarray}
  \frac{d^3\Gamma_H}{dm_\star d\cos\theta d\cos\vartheta}
\!&=&\! 
\frac{d\Gamma_H}{dm_\star}\,\frac{9}{16}\,
      \bigg\{ \frac{4m_\star^2 m_Z^2+m_H^4\kappa^2}{12 m^2_Z m^2_\star + m^4_H \kappa^2}\,
                  \sin^2\theta\sin^2\vartheta 
                  \nonumber\\
&&\!\!\!\!\!  + \frac{2m^2_\star m^2_Z}{12 m^2_Z m^2_\star + m^4_H \kappa^2}\,
                   [(1+\cos^2\theta)(1+\cos^2\vartheta) 
                        +4\eta_\ell \eta_f \cos\theta\cos\vartheta]\bigg\},
\label{eq:correlated_invariant-mass_polar_angles_distribution}
\end{eqnarray}
with the invariant-mass distribution in 
Eq.$\,$(\ref{eq:sm_higgs_invariant_mass_distribution}), 
while the corresponding distribution with respect to the azimuthal angle
$\varphi$ is
\begin{eqnarray}
      \frac{d^2\Gamma_H}{dm_\star d\varphi}
= \frac{d\Gamma_H}{dm_\star}\, 
        \frac{1}{2\pi}\bigg[ 1 -\eta_\ell \eta_f \frac{9\pi^2}{16}
       \frac{m_Z m_\star(m^2_H-m^2_Z-m^2_\star)}{ 
             12m^2_Z m^2_\star + m^4_H \kappa^2} \,
            \cos\varphi 
   + \frac{2 m^2_Z m^2_\star }{12 m^2_Z m^2_\star +m^4_H\kappa^2} 
      \cos 2\varphi \bigg],
\label{eq:correlated_invariant-mass_azimuthal_angle_distribution}
\end{eqnarray}
where $\eta_{\ell}=\eta_f=(v^2_{-} - v^2_{+})/(v^2_{-}+v^2_{+})$ 
is the polarization asymmetry factor of the fermion in the decay
processes $Z\rightarrow \ell^-\ell^+$ and $Z\rightarrow f\bar{f}$ with  
$\ell, f=e, \mu$, and $\gamma_{\star, Z}$ and $\beta_{\star, Z}$ are 
the boost factors and speeds of the off-shell $Z^\star$ and  
on-shell $Z$ bosons in Eq.$\,$\eqref{eq:boost_factors}, 
respectively, in the $H$ rest frame. Numerically, the polarization
asymmetry factor $\eta_{\ell,f}=0.237$ for a charged lepton 
of $f=e, \mu$.  After integrating the correlated 
invariant-mass and polar-angle distribution 
\eqref{eq:correlated_invariant-mass_polar_angles_distribution}  
over the invariant mass $m_\star$ from $0$ to $m_H-m_Z$, we
obtain the normalized polar-angle correlation as
\begin{eqnarray}
  \frac{d^2\mathcal{N}_H}{ d\cos\theta d\cos\vartheta}
= 
\frac{9}{16}\!
      \bigg\{ \frac{R_{L}(\omega_Z)}{R(\omega_Z)}\!
                  \sin^2\theta\sin^2\vartheta 
\!+\! \frac{R_{T}(\omega_Z)}{R(\omega_Z)}\!
                   \bigg[
                   \frac{(1\!+\!\cos^2\theta)(1\!+\!\cos^2\vartheta)}{4} 
                \! +\!\eta_\ell \eta_f \cos\theta\cos\vartheta\bigg]\bigg\},
\label{eq:sm_higgs_polar-angle_correlated_distribution}
\end{eqnarray}
of which the integral over the polar angle $\vartheta$ leads 
to the normalized  single polar-angle distribution 
\begin{eqnarray}
 \frac{d{\cal N}_H}{d\cos\theta}
= \frac{3}{4}\, 
  \left[\frac{R_{L}(\omega_Z)}{R(\omega_Z)}\, \sin^2\theta 
       +\frac{R_{T}(\omega_Z)}{R(\omega_Z)}\, \frac{1+\cos^2\theta}{2}\right],
\end{eqnarray}
with the $\omega_Z$ dependent functions given by
\begin{eqnarray}
  R_{L}(\omega_Z)
\!&=&\! (1-2\omega_Z^2)^2 A_0(\omega_Z)
        -2(1-2\omega_Z^2)A_1(\omega_Z)
        +A_2(\omega_Z),\\
  R_{T}(\omega_Z)
\!&=&\! 8\omega_Z^2\big[\omega_Z^2 A_0(\omega_Z)+A_1(\omega_Z)\big],
\end{eqnarray}
with $R(\omega_Z)=R_{L}(\omega_Z)+R_{T}(\omega_Z)$ identical 
to the expression (\ref{eq:function_r_x}). 
Integrating the correlated distribution 
\eqref{eq:correlated_invariant-mass_azimuthal_angle_distribution}
over the invariant mass yields the azimuthal-angle distribution
\begin{eqnarray}
  \frac{d{\cal N}_H}{d\varphi}
= \frac{1}{2\pi}\, 
   \left[1
         +\eta_\ell \eta_f\frac{9\pi^2}{64} 
          \frac{R_{1}(\omega_Z)}{R(\omega_Z)}\cos\varphi
         + \frac{R_{2}(\omega_Z)}{R(\omega_Z)}\cos 2\varphi
  \right],
\end{eqnarray}
where the explicit expressions of the $\omega_Z$-dependent functions,
$R_1$ and $R_2$, can be decomposed  as
\begin{eqnarray}
  R_{1}(\omega_Z)
&=& -4\omega_Z\big[ (1-2\omega^2_Z) B_0(\omega_Z)
  - B_1(\omega_Z)\big],
  \\
  R_{2}(\omega_Z)
&=& R_{T}(\omega_Z)/4,
\end{eqnarray}
where the integral-form definition of the $\omega_Z$-dependent
functions $A_{n}$, $B_{n}$ and their explicit forms expressed in terms of 
three types of elliptic integrals can be found in Appendix~\ref{appedix:integral_functions}.
Quantitatively, for $m_Z=91\, {\rm GeV}$ and $m_H=125\, {\rm GeV}$,  
$R_{L} = 0.59 R$, $R_{T}=0.41 R$, $R_{1}=-0.64\,R$ and $R_2=0.10\, R$.\s

\setcounter{equation}{0}

\section{Vetoing Higgs imposters}
\label{sec:vetoing_higgs_imposters}

In this section, we investigate step by step how to veto all 
the Higgs imposters unambiguously through the invariant-mass threshold behavior 
and 4-lepton polar and azimuthal angle distributions by identifying
the features of each Higgs imposter distinct from those of the SM
Higgs boson. \s

\subsection{The case with a cascade decay 
            $H\to Z^\star Z \to \ell^-\ell^+ Z$}
\label{subsec:only_j0=1_terms}

First, we consider the case when all the contact terms are 
negligible and the $\ell^-\ell^+$ final state is generated simply 
through an off-shell vector boson $Z^\star$ fixing the value 
of $j_0$ to be 1 in the CC case. In general, a photon-mediated part 
might be present. However, this contribution can be easily ruled out 
as its photon propagator part leads to a significantly different 
invariant-mass distribution from the $Z$ propagator part, 
especially in the region with a small invariant mass $m_\star$ due to 
the huge enhancement factor of $m^4_Z/m^4_\star$. 
The helicity amplitude of the decay $H\to Z_\star Z \to \ell^-\ell^+ Z$ 
can be decomposed into two parts as
\begin{eqnarray}
   {\cal M}^{[H\to\ell^-\ell^+ Z]}_{\lambda_H,\sigma,\lambda_Z}
= 8 G_F m^2_Z  {\cal T}_{\lambda_\star,\lambda_Z}\,
   \frac{m_Z m_\star}{m^2_\star-m^2_Z+im_Z\Gamma_Z}\,
   v_\sigma\, 
   d^1_{\lambda_\star, \sigma}(\theta),
\end{eqnarray}
with the constraint $\lambda_\star=\lambda_H+\lambda_Z=\pm 1, 0$ for 
the helicity of the off-shell $Z^\star$ boson,  $\sigma=\pm 1$,
and the chiral vector couplings, $v_\sigma$, in the CC case.
Here, the reduced helicity amplitudes ${\cal T}_{\lambda_\star,\lambda_Z}$ 
of the two-body decay $H\to Z^\star Z$ can be
re-written apart from a constant coefficient as
\begin{eqnarray}
  {\cal T}_{\lambda_\star,\lambda_Z}
= \epsilon^{*\beta}_Z(k_Z,\lambda_Z)
  \epsilon^{*\mu}_\star(k,\lambda_\star)\,
  [{\cal H}^{1[HZ]}_{[\lambda_H,\lambda_Z]}]_{\bar{\alpha},\beta,\mu}
  \epsilon^{\bar{\alpha}}(p_H,\lambda_H),
\end{eqnarray}
in terms of the helicity-specific tensor 
$[{\cal H}^{1[HZ]}_{[\lambda_H,\lambda_Z]}]$ with the constraint 
$\lambda_\star=\lambda_H+\lambda_Z = \pm 1, 0$
for any value of the $H$ spin, because all the polarization axes of 
the $H$, $Z_\star$ and $Z$ bosons are set to be along the $z$ axis.
It is noteworthy that any general $HZ^\star Z$ vertex 
can be expressed effectively as a linear combination of the basic 
operators $[{\cal H}^{1[HZ]}]$, of which the independent number
is three (3) for $s_H=0$, seven (7) for $s_H=1$ and nine (9) for 
$s_H\geq 2$ with no more increase of the number of independent terms, 
as analyzed
systematically in Sec.$\,$\ref{sec:weaving_covariant_4-point_vertices}.\s

\subsubsection{Fully-correlated $CP$-invariant distributions}
\label{subsubsec:fully-correlated_distribution}

In a $CP$-invariant theory, the $H$-boson state can be attributed to 
a conserved quantum number called normality $n_H = (-1)^{s_H} P$, which 
links the helicity amplitudes that are connected under parity 
transformations. If the interactions determining 
the $HZ^\star Z$ vertex are parity-invariant, the reduced helicity amplitudes 
are related as,
\begin{eqnarray}
  {\cal T}_{\lambda_\star, \lambda_Z}
= n_H \, {\cal T}_{-\lambda_\star, -\lambda_Z},
\end{eqnarray}
with the $H$ normality, $n_H=(-1)^{s_H} P$. The fully-combined invariant-mass
and angular correlation is decomposed into two parts\footnote{
Although the case with mixed normality states is not investigated in the 
present work, we note that it is straightforward to extend
our formalism to accommodate the mixed case. }
\begin{eqnarray}
     \frac{d^4\Gamma_H}{dm_\star d\cos\theta\cos\vartheta d\varphi}
\sim \frac{m^3_\star\,\kappa}{(m^2_\star-m^2_Z)^2+m^2_Z\Gamma^2_Z}\,
     \frac{d^3{\Gamma}_H}{d\cos\theta d\cos\vartheta d\varphi},
\label{eq:fully_correlated_distribution_with_virtual_z}
\end{eqnarray}
apart from a constant coupling factor, where the full angular correlation part 
is given  by
\begin{eqnarray}
  \frac{d^3{\Gamma_H}}{d\cos\theta d\cos\vartheta d\varphi}
&\sim & 2\sin^2\theta\sin^2\vartheta\, |{\cal T}_{0,0}|^2
      + (1+\cos^2\theta)(1+\cos^2\vartheta)\,
       [|{\cal T}_{+1, +1}|^2 +|{\cal T}_{+1,-1}|^2] \nonumber\\
&& + 2(1+\cos^2\theta)\sin^2\vartheta\, |{\cal T}_{+1, 0}|^2
   + 2\sin^2\theta (1+\cos^2\vartheta)\, |{\cal T}_{0, +1}|^2 \nonumber\\
&& + 4 \sin\theta\cos\theta\sin\vartheta\cos\vartheta\cos\varphi\,
     {\rm Re}({\cal T}_{+1,+1}{\cal T}^*_{0,0}
             -{\cal T}_{+1,0}{\cal T}^*_{0,-1})\nonumber\\
&& + \sin^2\theta\sin^2\vartheta\cos2\varphi\, 
    {\rm Re}({\cal T}_{+1,+1} {\cal T}^*_{-1,-1}) \nonumber\\
&& -4\eta_{\ell} \sin\theta\sin\vartheta\cos\vartheta\sin\varphi\,
   {\rm Im}({\cal T}_{+1,+1}{\cal T}^*_{0,0}
           -{\cal T}_{+1,0}{\cal T}^*_{0,-1})\nonumber\\
&& -4\eta_{f} \sin\theta\cos\theta\sin\vartheta\sin\varphi\,
   {\rm Im}({\cal T}_{+1,+1}{\cal T}^*_{0,0}
           +{\cal T}_{+1,0}{\cal T}^*_{0,-1})\nonumber\\
&& +4\eta_{\ell}\eta_f \cos\theta\cos\vartheta\,
      [|{\cal T}_{+1,+1}|^2-|{\cal T}_{+1,-1}|^2]\nonumber\\
&& +4\eta_{\ell}\eta_f \sin\theta\sin\vartheta\cos\varphi\,
     {\rm Re}({\cal T}_{+1,+1}{\cal T}^*_{0,0}
             +{\cal T}_{+1,0}{\cal T}^*_{0,-1}).
\end{eqnarray}
in terms of the reduced helicity amplitudes 
${\cal T}_{\lambda_\star,\lambda_Z}$ describing the two-body
decay $H\to Z^\star Z$. Integrating the correlated distribution only over 
the azimuthal angle $\varphi$ generates a correlated polar-angle
distribution
\begin{eqnarray}
  \frac{d^2{\Gamma_H}}{d\cos\theta d\cos\vartheta}
&\sim & 2\sin^2\theta\sin^2\vartheta\, |{\cal T}_{0,0}|^2
      + (1+\cos^2\theta)(1+\cos^2\vartheta)\,
       [|{\cal T}_{+1, +1}|^2 +|{\cal T}_{+1,-1}|^2] \nonumber\\
&& + 2(1+\cos^2\theta)\sin^2\vartheta\, |{\cal T}_{+1, 0}|^2
   + 2\sin^2\theta (1+\cos^2\vartheta)\, |{\cal T}_{0, +1}|^2 \nonumber\\
&& +4\eta_{\ell}\eta_f \cos\theta\cos\vartheta\,
      [|{\cal T}_{+1,+1}|^2-|{
      \cal T}_{+1,-1}|^2],
\end{eqnarray}
while integrating the correlated distribution over
the polar and azimuthal angles, $\vartheta$ and $\varphi$, generates 
a single polar-angle distribution
\begin{eqnarray}
 \frac{d{\Gamma_H}}{d\cos\theta}
& \sim &   \sin^2\theta\, 
(|{\cal T}_{0,0}|^2+2|{\cal T}_{0, +1}|^2)
      + (1+\cos^2\theta)\,
       (|{\cal T}_{+1, +1}|^2 +|{\cal T}_{+1,-1}|^2
       +|{\cal T}_{+1, 0}|^2),
\end{eqnarray}
which is to be combined with $Z^\star$ propagator factor as in 
Eq.$\,$(\ref{eq:fully_correlated_distribution_with_virtual_z}) before 
being utilized for characterizing the spin and parity of the $H$ boson
explicitly in the following. On the other hand, integrating the correlated 
distribution over the two polar angles, $\theta$ and $\vartheta$, 
generates a single azimuthal-angle distribution 
\begin{eqnarray}
     \frac{d{\Gamma_H}}{d\varphi}
\sim \sum_{\lambda_\star, \lambda_Z}
       |{\cal T}_{\lambda_\star, \lambda_Z}|^2 
       +\frac{9\pi^2}{32} \eta_{\ell}\eta_f \cos\varphi\,
     {\rm Re}({\cal T}_{+1,+1}{\cal T}^*_{0,0}
             +{\cal T}_{+1,0}{\cal T}^*_{0,-1})
       +
       \frac{1}{2}\cos2\varphi\, 
    {\rm Re}({\cal T}_{+1,+1} {\cal T}^*_{-1,-1}) ,  
\end{eqnarray}
which is to be combined with $Z^\star$ propagator factor as in 
Eq.$\,$(\ref{eq:fully_correlated_distribution_with_virtual_z}) for
the correlated $m_\star$ and $\varphi$ distribution.\s

\subsubsection{Ruling out all the Higgs imposters}
\label{subsubsec:ruliing-out_higgs_imposters}

The leading $\kappa$ dependence of the helicity amplitudes can be determined
by counting the number of momenta in each term of the tensor 
$[{\cal H}^{1[HZ]}_{[\lambda_H,\lambda_Z]}]_{\bar{\alpha},\beta,\mu}$.
Each re-scaled momentum, which is represented by $\hat{p}_{Z\beta}$ 
and/or $\hat{r}_{H\alpha}$, contracted with the $Z$-boson polarization 
vector or the $H$-boson polarization tensor will necessarily give zero or one power 
of $\kappa$. Furthermore, the normalized momentum $\hat{p}_{\star\mu}$ 
contracted with the lepton current will give rise to one power of 
$\kappa$ due to  the transversality of the current. 
The overall $\kappa$ dependence of the
invariant mass spectrum can be derived from the absolute square of the
helicity amplitude multiplied by a single factor $\kappa$ from 
the phase space.\s

\begin{itemize}
\item {\bf Spin 0.} 
If the particle $H$ is an odd-normality pseudo-scalar, i.e. its parity is odd,
the $HZ^\star Z$ vertex term is given by a parity-odd tensor  
in any CP-invariant theory  which is expressed by a linear combination of 
the basic $HZ$ operators for the spin-0 $H$ boson 
in Eq.~\eqref{eq:spin-0_HZ_operators} as
\begin{eqnarray}
 \frac{1}{m_H^2}\langle \mu\beta pq\rangle
=  \kappa\, \langle \mu \beta\hat{p}\hat{r}\rangle
= -i  \kappa\, (U^+_{\mu\beta}-U^-_{\mu\beta}),
\end{eqnarray}
with $p=k+k_Z$ and $q=k-k_Z$,  giving rise to ${\cal T}_{0,0}=0$
and non-zero ${\cal T}_{+1,+1}= -{\cal T}_{-1,-1}\propto i \kappa$ apart
from a constant coefficient.
Consequently, the invariant mass spectrum decreases in proportion to
$\kappa^3$ near the kinematical limit and the single polar-angle 
distribution is simply proportional to $(1+\cos^2\theta)$ with 
no $\sin^2\theta$ term. Explicitly, the normalized invariant-mass
distribution is given by
\begin{eqnarray}
  \frac{d{\cal N}_H}{dm_\star}
= \frac{m_\star^3\kappa^2}{\mathcal{P}(\omega_Z)
  \big[(m_\star^2-m_Z^2)^2+m_Z^2\Gamma_Z^2\big]}\, \kappa, 
\end{eqnarray}
 with the $\omega_Z$-dependent function ${\cal P}$ decomposed as
\begin{eqnarray}
 \mathcal{P}(\omega_Z)
= \omega_Z^2(1-4\omega^2_Z)\, A_0(\omega_Z)
+(1-6\omega^2_Z)\, A_1(\omega_Z)
-(2-\omega^2_Z)\, A_2(\omega_Z)
+A_3(\omega_Z),  
\end{eqnarray}
where the explicit form of the $\omega_Z$-dependent function $A_n(\omega_Z)$ 
with $n$ from 0 to 5
in the limit of neglecting the $Z$-boson decay width $\Gamma_Z$ is 
listed in Appendix~\ref{appedix:integral_functions}.
The normalized single  polar-angle and azimuthal-angle distributions 
are given simply by
\begin{eqnarray}
\frac{d{\cal N}_H}{d\cos\theta}
= \frac{3}{8} (1+\cos^2\theta) \quad \mbox{and}\quad
   \frac{d{\cal N}_H}{d\varphi}
= \frac{1}{2\pi} \left(1-\frac{1}{4}\cos 2\varphi\right),
\end{eqnarray}
which are completely independent of both the $H$ and $Z$-boson masses.
Therefore, it is definite that the spin-0 and parity-even SM Higgs boson 
cannot be imitated by any pseudo-scalar Higgs imposter. 
For various previous detailed studies 
on the general scalar $H$ couplings, see for example 
Refs.~\cite{Hohlfeld:2000tpa,Kramer:1993jn,Hagiwara:1993sw,
Hagiwara:2000tk,Grzadkowski:2000hm,Han:2000mi}.
\item {\bf Spin 1.} Every term in
$[{\cal H}^{1[HZ]}_{[\lambda_H,\lambda_Z]}]_{\bar{\alpha},\beta,\mu}$ 
involves at least one power of momentum. This seems to imply that every 
helicity amplitude vanishes near the threshold linearly in 
$\kappa$. However, there arises a critical issue that requires a solution 
by incorporating a suitable factor dependent on the parameter $\kappa$. 
In general, the spin-1 tensor may include an odd-normality term
$\langle \alpha\beta\mu_1 \hat{p}\rangle$.  This tensor is decomposed 
effectively into the $HZ$ operators  with $s_H=1$ 
in Table.~\ref{tab:spin-1_covariant_operators} as 
\begin{eqnarray}
  \langle \mu\beta\alpha\hat{p}\rangle
= \frac{ie_Z}{2\omega_Z\kappa} (U^+_{\mu\alpha}-U^-_{\mu\alpha})\hat{p}_{Z\beta}
 +\frac{ie_\star}{2\omega_\star\kappa} (U^+_{\beta\alpha}-U^-_{\beta\alpha})\hat{p}_{\star\mu}
  -\frac{i}{\kappa}(U^+_{\mu\beta}-U^-_{\mu\beta}) \hat{r}_{H\alpha}.
\end{eqnarray}
As a result of this decomposition, we can observe straightforwardly 
that the helicity amplitudes do not vanish at the threshold, even after 
the operator is contracted with the $H$, $Z^\star$ 
and $Z$ polarization vectors. On the other hand, this decomposition leads to 
${\cal T}_{0,0}=0$ so that no $\sin^2\theta \sin^2\vartheta$ polar-angle 
correlation is generated.
Instead, it gives rise to the $\sin^2\theta(1+\cos^2\vartheta)$ 
and $(1+\cos^2\theta)\sin^2\vartheta$ polar-angle
correlations, which are not present in the SM.
Explicitly, the normalized invariant-mass distribution reads
\begin{eqnarray}
  \frac{d\mathcal{N}_{H}}{dm_\star}
= \frac{m_\star [12m_\star^2m_Z^2
    + m^2_H (m_\star^2+m_Z^2)\kappa^2]}{
         m_H^2{\cal K}(\omega_Z) 
      [ (m_\star^2-m_Z^2)^2+m_Z^2\Gamma_Z^2]}\, \kappa, 
\end{eqnarray}
where the $\omega_Z$-dependent function $\mathcal{K}$ is given by
\begin{eqnarray}
  \mathcal{K}(\omega_Z)
= 2\omega_Z^2(1+2\omega_Z^2) \,A_0(\omega_Z)
+(1+4\omega_Z^2) \,A_1(\omega_Z)
-2(1-\omega_Z^2) \,A_2(\omega_Z)
+A_3(\omega_Z).
\end{eqnarray}
The normalized single polar-angle distribution can be decomposed into two
parts as
\begin{eqnarray}
  \frac{d\mathcal{N}_{H}}{d\cos\theta}
= \frac{3}{4}
  \bigg[
  \frac{\mathcal{K}_L(\omega_Z)}{\mathcal{K}(\omega_Z)}
  \sin^2\theta
  +\frac{\mathcal{K}_T(\omega_Z)}{\mathcal{K}(\omega_Z)}
  \frac{(1+\cos^2\theta)}{2}\bigg].
\end{eqnarray}
The expressions of two $\omega_Z$-dependent functions $\mathcal{K}_{T, L}$ 
are given by
\begin{eqnarray}
    \mathcal{K}_L(\omega_Z)
&=& \omega_Z^2\big[A_0(\omega_Z)+2A_1(\omega_Z)+A_2(\omega_Z)\big], \\
    \mathcal{K}_T(\omega_Z)
&=& \omega_Z^2(1+4\omega_Z^2)\,A_0(\omega_Z)
+(1+2\omega_Z^2)\,A_1(\omega_Z) \nonumber\\
&& -(2-\omega_Z^2)\,A_2(\omega_Z)
+A_3(\omega_Z),
\end{eqnarray}
of which the sum is $\mathcal{K}=\mathcal{K}_L+\mathcal{K}_T$.
Numerically, 
${\cal K}_L= 0.498\, {\cal K}$ and ${\cal K}_T= 0.502\, {\cal K}$ 
for $m_Z=91$ GeV and $m_H=125$ GeV.
And the normalized azimuthal-angle distribution by
\begin{eqnarray}
  \frac{d\mathcal{N}_{H}}{d\varphi}
= \frac{1}{2\pi}
  \bigg[1+\frac{9\pi^2}{64}
  \eta_\ell \eta_f 
  \frac{\mathcal{K}_1(\omega_Z)}{\mathcal{K}(\omega_Z)}\cos\varphi
   +\frac{\mathcal{K}_2(\omega_Z)}{\mathcal{K}(\omega_Z)} 
   \cos2\varphi
   \bigg],
\end{eqnarray}
where the expressions of two $\omega_Z$-dependent functions 
$\mathcal{K}_{1,2}$ are 
\begin{eqnarray}
    \mathcal{K}_1(\omega_Z)
&=& -\omega_Z\big[B_0(\omega_Z)-B_2(\omega_Z)\big],
\\
    \mathcal{K}_2(\omega_Z)
&=& -\omega_Z^2
\big[\omega_Z^2A_0(\omega_Z)+A_1(\omega_Z)\big].
\end{eqnarray}
Numerically, ${\cal K}_1= -0.34\, {\cal K}$ and 
${\cal K}_2= -0.06\, {\cal K}$ for $m_Z=91$ GeV and $m_H=125$ GeV.
The integral-form definition and integrated expression of each $A_n(\omega_Z)$ 
with $n=0$ to 5  and those of each $B_n(\omega_Z)$ from $n=0$ to 4
are listed in Appendix~\ref{appedix:integral_functions}.
\item {\bf Spin 2.} The general $HZ^\star Z$ vertex for the spin-2 $H$ 
contains a term with no momentum dependence
\begin{eqnarray}
  [{\cal H}^{1[HZ]}]_{\alpha_1\alpha_2, \mu, \beta}
\, \sim\,  g_{\alpha_1\mu} g_{\alpha_2\beta},
\end{eqnarray}
generating helicity amplitudes that do not vanish at the threshold.
As a concrete spin-2 example, we adopt the interaction
vertex of the even-normality first Kaluza-Klein (KK) graviton with 
a virtual $Z^\star$ boson 
and an on-shell $Z$ boson in the framework of large extra 
dimensions~\cite{Arkani-Hamed:1998jmv, Antoniadis:1998ig}.
Apart from a gauge-fixing term with no
physical impact, the covariant  $HZ^\star Z$ vertex for the spin-2 $H$ is 
given by
\begin{eqnarray}
   \Gamma_{\mu\beta;\alpha_1\alpha_2}
= \frac{2}{m_H}\left[ (m^2_Z+k\cdot k_Z) g_{\mu\alpha_1} g_{\beta\alpha_2}
                    -g_{\mu\alpha_1}k_\beta k_{Z\alpha_2} 
                    -g_{\beta\alpha_1}k_{Z\mu} k_{\alpha_2}  
                    +g_{\mu\beta} k_{\alpha_1}k_{Z\alpha_2}\right],
\end{eqnarray}
up to an overall coupling constant,
where the indices $\alpha_1\alpha_2$ are for the symmetric and traceless
spin-2 wave tensor of the decaying particle $H$ and the indices $\mu$ 
and $\beta$ are for the
virtual $Z^\star $ and real $Z$ polarization  
vectors~\cite{Han:1998sg}. 
The vertex of the spin-2 $H$ imposter can be expressed in terms of
the general $HZ$ operators $[{\cal H}^{1[HZ]}_{[\lambda_H,\lambda_Z]}]$ 
in Table.~\ref{tab:spin-2_covariant_operators}
effectively as
\begin{eqnarray}
   \Gamma_{\mu\beta;\alpha_1\alpha_2}  
&=& m_H\bigg[\, e_Z\, (U^+_{\mu\alpha_1}U^+_{\beta\alpha_2}  
                     +U^-_{\mu\alpha_1}U^-_{\beta\alpha_2}) 
     + \frac{e_Z-\kappa^2}{2\kappa^2}\, 
            (U^+_{\mu\beta}+U^-_{\mu\beta})\,
            \hat{r}_{H\alpha_1}\hat{r}_{H\alpha_2}
\nonumber \\
&& \qquad -\frac{e^2_Z-\kappa^2}{2\omega_Z\kappa^2} \,
          (U^+_{\mu\alpha_1} + U^-_{\mu\alpha_1})\,
           \hat{p}_{Z \beta} \hat{r}_{H\alpha_2}
         + \frac{e_\star e_Z-\kappa^2}{2\omega_\star \kappa^2}
           (U^+_{\beta\alpha_1} + U^-_{\beta\alpha_1})\,
           \hat{p}_{\star \mu} \hat{r}_{H\alpha_2} 
\nonumber\\
&& \qquad - \frac{e_\star e^2_Z-\kappa^2(1+\omega^2_\star+\omega^2_Z)}{
                  4\omega_\star \omega_Z \kappa^4}\,
            \hat{p}_{\star \mu} \hat{p}_{Z \beta}
            \hat{r}_{H \alpha_1}\hat{r}_{H\alpha_2}\bigg].
\end{eqnarray} 
The vertex terms contributing to the reduced helicity amplitudes, ${\cal T}_{\pm 1, 0}$ and ${\cal T}_{0, \pm 1}$, lead to
non-trivial $(1+\cos^2\theta)\sin^2\vartheta$ and $\sin^2\theta(1+\cos^2\vartheta)$ 
correlations, respectively, which are absent in the SM.
Therefore, even though the invariant mass spectrum decreases linearly 
in $\kappa$, these non-trivial polar-angle correlations enable us 
to rule out effectively the spin-2 KK-graviton imposter assignment 
to the state. 
Explicitly, the normalized invariant-mass distribution is given by
\begin{align}
\frac{d\mathcal{N}_H}{dm_\star}
=
\frac{m_\star\big[120m_\star^2m_Z^2
-10m_H^2(m_H^2-3m_\star^2-2m_Z^2)\kappa^2
+m_H^2(10m_H^2+2m_\star^2+m_Z^2)\kappa^4\big]}{m_H^2\mathcal{G}(\omega_Z)[(m_\star^2-m_Z^2)^2+m_Z^2\Gamma_Z^2]}  \kappa, 
\end{align}
where the $\omega_Z$-dependent function $\mathcal{G}$ is given by
\begin{eqnarray}
  \mathcal{G}(\omega_Z)
&=& \omega_Z^2(13+56\omega_Z^2+48\omega_Z^4) A_0(\omega_Z)
+4(3+8\omega_Z^2+20\omega_Z^4)A_1(\omega_Z)
\nonumber
\\
&&
-2(9-10\omega_Z^2+12\omega_Z^4) A_2(\omega_Z)
+2(1-14\omega_Z^2)A_3(\omega_Z)
\nonumber
\\
&&
+(2+3\omega_Z^2) \,A_4(\omega_Z)+2A_5(\omega_Z).
\end{eqnarray}
Integrating the fully-correlated distribution over the invariant mass
leads to the polar-angle correlation as 
\begin{eqnarray}
  \frac{d^2\mathcal{N}_H}{ d\cos\theta d\cos\vartheta}
\!&=&\! 
\frac{9}{16}\,
      \bigg\{ \frac{\mathcal{G}_{LL}(\omega_Z)}{\mathcal{G}(\omega_Z)}\,\sin^2\theta\sin^2\vartheta 
      + \frac{\mathcal{G}_{TT}(\omega_Z)}{\mathcal{G}(\omega_Z)}
      \frac{(1+\cos^2\theta)(1+\cos^2\vartheta)}{4}
                  \nonumber\\
&&\!\!\!\!\!  
+\frac{\mathcal{G}_{TL}(\omega_Z)}{\mathcal{G}(\omega_Z)}\,\frac{(1+\cos^2\theta)}{2}\sin^2\vartheta
+\frac{\mathcal{G}_{LT}(\omega_Z)}{\mathcal{G}(\omega_Z)}\,\sin^2\theta\frac{(1+\cos^2\vartheta)}{2}
                   \nonumber\\
&&\!\!\!\!\!  
     +\eta_\ell \eta_f  \frac{\mathcal{G}'_{TT}(\omega_Z)}{\mathcal{G}(\omega_Z)}\cos\theta\cos\vartheta\bigg\},
\label{eq:spin-2_polar-angle_correlated__distribution}
\end{eqnarray}
where the five $\omega_Z$-dependent ${\cal G}$ functions are given by
\begin{eqnarray}
    \mathcal{G}_{LL}(\omega_Z)
&=& \omega_Z^2\Big[(1+8\omega_Z^2+16\omega_Z^4)A_0(\omega_Z)
+8(1+4\omega_Z^2)\,A_1(\omega_Z)
\nonumber\\
&& +2(7-4\omega_Z^2)\,A_2(\omega_Z)
-8A_3(\omega_Z)+A_4(\omega_Z)\Big],
\\
    \mathcal{G}_{TT}(\omega_Z)
&=& 2\Big[2\omega_Z^2(3+8\omega_Z^4)A_0(\omega_Z)
+6(1-2\omega_Z^2+4\omega_Z^4) A_1(\omega_Z)
\nonumber\\
&&  -(12-15\omega_Z^2+8\omega_Z^4) A_2(\omega_Z)
+(7-10\omega_Z^2)A_3(\omega_Z)
\nonumber\\
&& -(2-\omega_Z^2)A_4(\omega_Z)
+A_5(\omega_Z)\Big],
\\
    \mathcal{G}_{TL}(\omega_Z)
&=& 24\omega_Z^2\Big[\omega_Z^2 A_0(\omega_Z)+
A_1(\omega_Z)\Big],
\\
    \mathcal{G}_{LT}(\omega_Z)
&=& 6\Big[4\omega_Z^4 A_0(\omega_Z)
+4\omega_Z^2 A_1(\omega_Z) +(1-4\omega_Z^2) A_2(\omega_Z)
\nonumber\\
&& 
-2A_3(\omega_Z)
+A_4(\omega_Z)\Big],
\\
    \mathcal{G}'_{TT}(\omega_Z)
&=& -2\Big[2\omega_Z^2(3-8\omega_Z^4)A_0(\omega_Z)
+6(1-2\omega_Z^2-4\omega_Z^4) A_1(\omega_Z)
\nonumber\\
&&  -(12+3\omega_Z^2-8\omega_Z^4) A_2(\omega_Z)
+5(1+2\omega_Z^2)A_3(\omega_Z)
\nonumber\\
&& +(2-\omega_Z^2)A_4(\omega_Z)
-A_5(\omega_Z)\Big],
\end{eqnarray}
satisfying $\mathcal{G}=\mathcal{G}_{LL}+\mathcal{G}_{LT}+\mathcal{G}_{TL}+\mathcal{G}_{TT}$.
Numerically, ${\cal G}_{LL}=0.18 \,{\cal G}$, 
${\cal G}_{LT}=0.25 \,{\cal G}$, ${\cal G}_{TL}=0.17 \,{\cal G}$,
${\cal G}_{TT}=0.4 \,{\cal G}$, and ${\cal G}'_{TT}=-0.3\,{\cal G}$.
Integrating the polar-angle correlation over the polar angle $\vartheta$ 
leads to the normalized single polar-angle distribution as
\begin{eqnarray}
  \frac{d\mathcal{N}_{H}}{d\cos\theta}
= \frac{3}{4}
  \bigg[
  \frac{\mathcal{G}_L(\omega_Z)}{\mathcal{G}(\omega_Z)}
  \sin^2\theta+
  \frac{\mathcal{G}_T(\omega_Z)}{\mathcal{G}(\omega_Z)}
  \frac{(1+\cos^2\theta)}{2}
   \bigg],
\end{eqnarray}
where the two $\omega_Z$-dependent functions $\mathcal{G}_{T, L}$ 
are given by
\begin{eqnarray}
    \mathcal{G}_L(\omega_Z)
&=& \omega_Z^2(1+32\omega_Z^2+16\omega_Z^4)A_0(\omega_Z)
+32\omega_Z^2(1+\omega_Z^2)\,A_1(\omega_Z)
\nonumber\\
&& +2(3-5\omega_Z^2-4\omega_Z^4)\,A_2(\omega_Z)
-4(3+2\omega_Z^2)A_3(\omega_Z)
\nonumber\\
&&+(6+\omega_Z^2)A_4(\omega_Z),
\\
    \mathcal{G}_T(\omega_Z)
&=& 2\Big[2\omega_Z^2(3+6\omega_Z^2+8\omega_Z^4)A_0(\omega_Z)
+6(1+4\omega_Z^4)\,A_1(\omega_Z)
\nonumber\\
&& \quad -(12-15\omega_Z^2+8\omega_Z^4)\,A_2(\omega_Z)
+(7-10\omega_Z^2)A_3(\omega_Z)
\nonumber\\
&&\quad -(2-\omega_Z^2)A_4(\omega_Z)
+A_5(\omega_Z)\Big],
\end{eqnarray}
of which the sum is $\mathcal{G}=\mathcal{G}_T+\mathcal{G}_L$.
Numerically, ${\cal G}_L= 0.43\, {\cal G}$ and 
${\cal G}_T= 0.57\, {\cal G}$ for $\omega_Z=m_Z/m_H=91/125=0.73$.
The normalized azimuthal-angle distribution is given by
\begin{eqnarray}
  \frac{d\mathcal{N}_{H}}{d\varphi}
= \frac{1}{2\pi}
  \bigg[1+\frac{9\pi^2}{64}
  \eta_\ell \eta_f 
  \frac{\mathcal{G}_1(\omega_Z)}{\mathcal{G}(\omega_Z)}\cos\varphi
   +\frac{\mathcal{G}_2(\omega_Z)}{\mathcal{G}(\omega_Z)} 
   \cos2\varphi\,
   \bigg],
\end{eqnarray}
where the expressions of two $\omega_Z$-dependent functions 
$\mathcal{K}_{1,2}$ are 
\begin{eqnarray}
    \mathcal{G}_1(\omega_Z)
&=& \omega_Z\big[16\omega_Z^2(1+\omega_Z^2)B_0(\omega_Z)
+(7+20\omega_Z^2)B_1(\omega_Z)
\nonumber\\
&&\quad \;\;-(3+8\omega_Z^2)B_2(\omega_Z)
-5B_3(\omega_Z)+B_4(\omega_Z)
\big],
\\
    \mathcal{G}_2(\omega_Z)
&=& \frac{\sqrt{6}}{2}
\big[4\omega_Z^4A_0(\omega_Z)
+\omega_Z^2(5-4\omega_Z^2)A_1(\omega_Z)
\nonumber\\
&&\quad\;\;\;
+(1-6\omega_Z^2)A_2(\omega_Z)
-(2-\omega_Z^2)A_3(\omega_Z)
+A_4(\omega_Z)
\big].
\end{eqnarray}
Numerically, ${\cal G}_1= 0.17\, {\cal G}$ and 
${\cal G}_2= 0.03\, {\cal G}$ for $\omega_Z=m_Z/m_H=91/125=0.73$.
The integral-form definition and integrated expression of each $A_n(\omega_Z)$ 
with $n=0$ to 5 are listed in Appendix~\ref{appedix:integral_functions}.
\item {\bf Spin \boldmath{$\geq 3$}.} The number of the general 
$HZ$ operators $[{\cal H}^{1[HZ]}_{[\lambda_H,\lambda_Z]}]$ 
with $j_0=1$
does not increase anymore. They are simply given by multiplying 
the general spin-2 $HZ^\star Z$ vertex
$\Gamma^{(2)}_{\alpha_1\alpha_2\mu\beta}$ by a $(s_H-2)$-th order 
product of the re-scaled momentum $\hat{r}_H=\kappa\hat{r}$ as
\begin{eqnarray}
  \Gamma^{(s_H)}_{\alpha_1\cdots\alpha_{s_H}\mu\beta}
= 
  \kappa^{s_H-2}\, [\Gamma^{(2)}_{\alpha_1\alpha_2\mu\beta}]\,
  \hat{r}_{\alpha_3}\cdots \hat{r}_{\alpha_{s_H}},
\end{eqnarray}
originating from the product $q_{\alpha_3}
\cdots q_{\alpha_{s_H}}$ with $q=k-k_Z = k_1+k_2-k_Z$. When contacted 
with the spin-$s_H$ wave tensor, the extra $s_H-2$ momenta give
rise to a leading power $\kappa^{s_H-2}$ in the helicity amplitudes.
The invariant mass spectrum therefore decreases near the threshold in
proportion to $\kappa^{2s_H-3}$, i.e., with a power 
$\geq 3$ for $s_H\geq 3$, 
in contrast to the single $\kappa$ power behavior of the SM Higgs boson.
\end{itemize}

\begin{figure}[ht!]
\vskip 0.4cm
\centering
\includegraphics[width=0.32\textwidth]{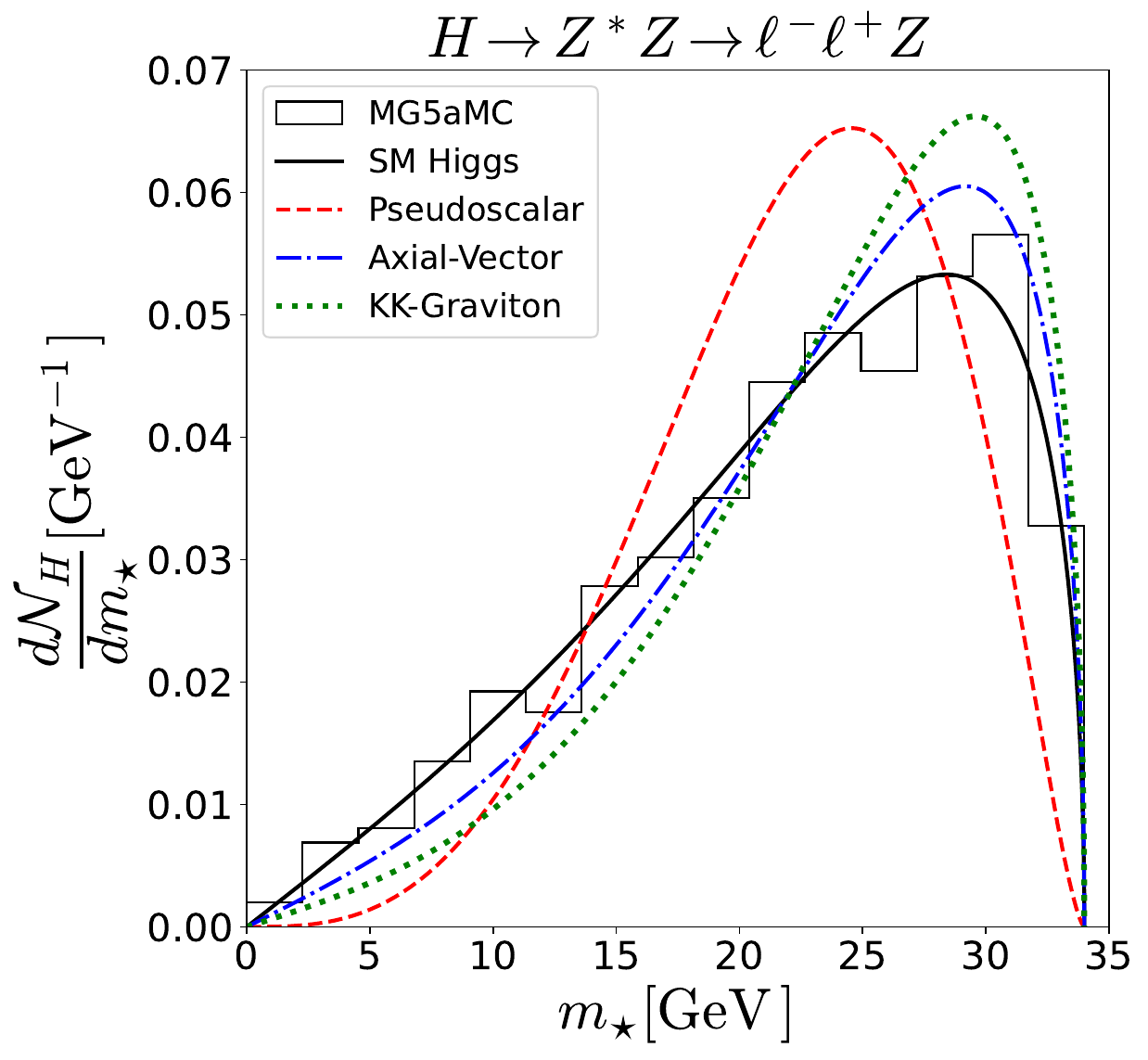}\,\,
\includegraphics[width=0.32\textwidth]{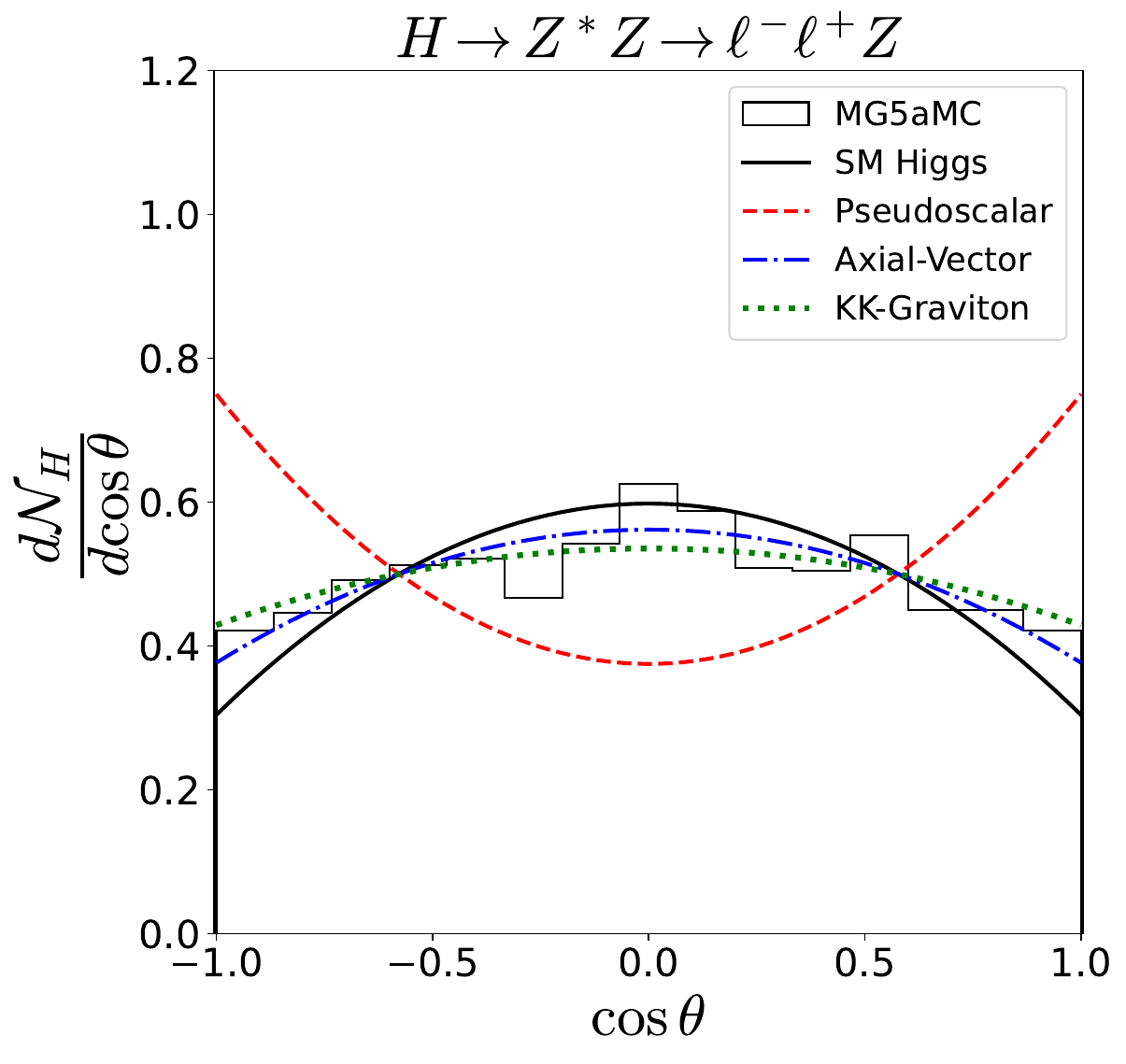}\,\, 
\includegraphics[width=0.32\textwidth]{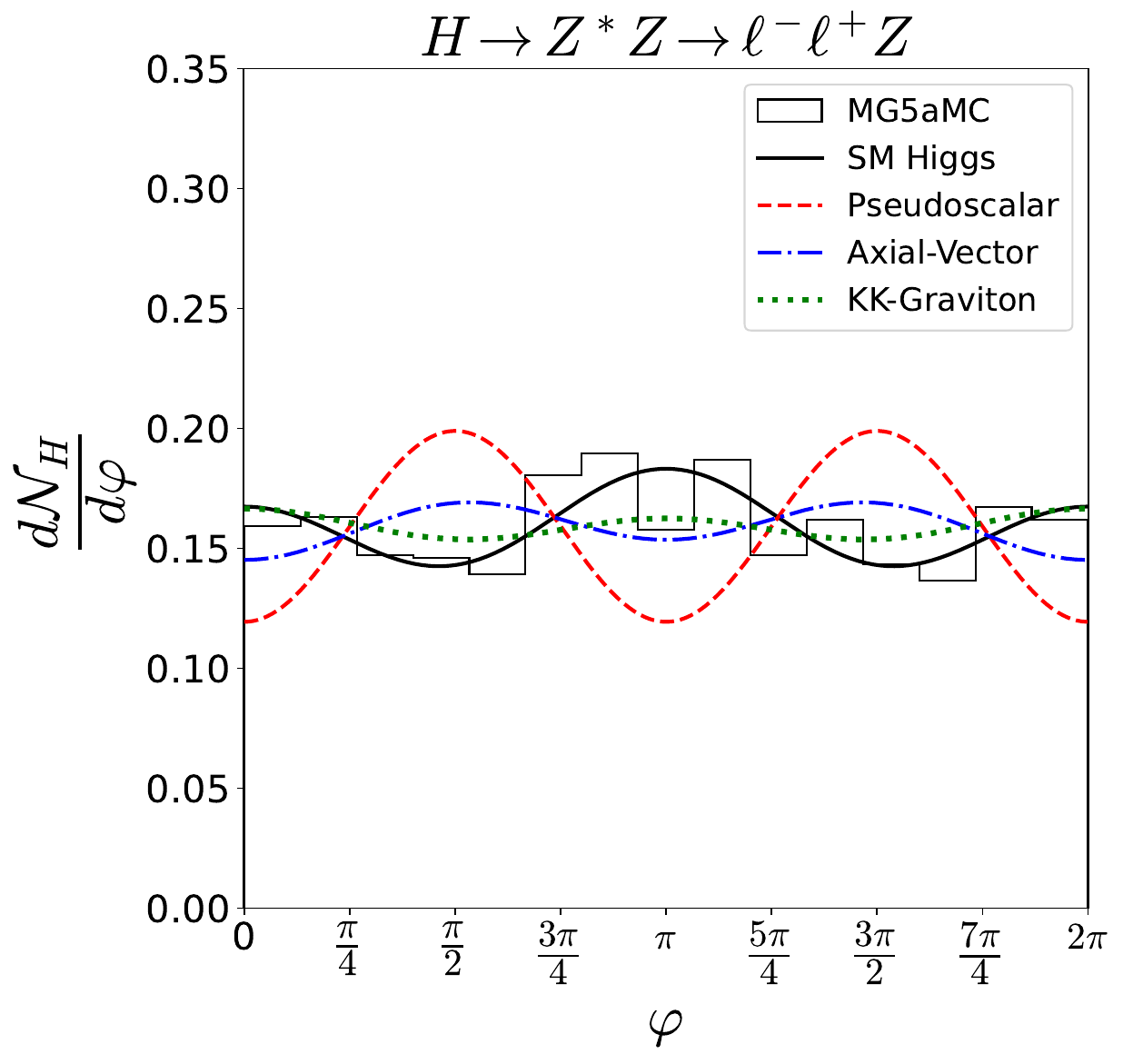}
\caption{\it Three characteristic distributions of 
   the SM Higgs boson (black solid line), a pseudoscalar 
   imposter (red dashed line),
   a spin-1 axial vector imposter (blue dot-dashed line), and a spin-2 KK 
   graviton imposter (green dotted line), respectively. 
   The left frame is the normalized invariant-mass distribution while
   the middle and right frames are the single polar-angle and 
   and azimuthal-angle distributions, respectively. The $Z$-boson 
   and $H$-boson masses are set to $m_Z=91\, {\rm GeV}$ and 
   $m_H=125\, {\rm GeV}$. 
   The histograms for the SM Higgs boson show the results from 
   about 1800 four-lepton events expected with essentially, 
   no kinematical cuts for an integrated luminosity of 300 ${\rm fb}^{-1}$ 
   at the LHC. 
   }
\label{fig:sm_higgs_invariant_mass_polar_angle_distribution}
\end{figure}

Reflecting all the detailed descriptions of the spin-0, spin-1, and spin-2 
examples for the cascade decay 
$H\rightarrow Z^\star Z \rightarrow \ell^-\ell^+ Z$ 
followed by the on-shell $Z$-boson decay $Z\rightarrow f\bar{f}$ with
$\ell, f=e, \mu$, we show in 
Fig.~\ref{fig:sm_higgs_invariant_mass_polar_angle_distribution} 
three characteristic distributions of the SM Higgs boson (black solid line), 
a pseudoscalar imposter (red dashed line), a spin-1 axial vector imposter 
(blue dot-dashed line) and a spin-2 KK  graviton imposter 
(green dotted line), respectively. The left frame is for the normalized
invariant-mass distribution 
while the middle and right frames are for the normalized single polar-angle
distribution and the normalized azimuthal-angle distributions, respectively. 
In this numerical illustration, the $Z$-boson and $H$-boson masses are set to 
$m_Z=91\, {\rm GeV}$ and $m_H=125\, {\rm GeV}$. The histograms for the SM 
Higgs boson shows the results from about 1800 events expected with essentially
no kinematical cuts for an integrated luminosity of 300 ${\rm fb}^{-1}$ at the 
LHC. For a real experimental measurement of the 4-lepton signal 
events we refer to Ref.~\cite{CMS:2021ugl}. 
The pseudoscalar
imposter can be clearly vetoed by each of the three distributions,
the spin-1 axial-vector imposter can be vetoed by the normalized 
azimuthal-angle distribution. 
In contrast, the mimicry of the SM Higgs boson by the spin-2 
KK graviton imposter in all three distributions poses a challenge 
in definitively distinguishing between the two. To achieve a clear
distinction, more sophisticated discriminatory distributions are 
required.\s

\begin{figure}[htb!]
\vskip 0.5cm
\begin{center}
\includegraphics[width=0.4\textwidth]{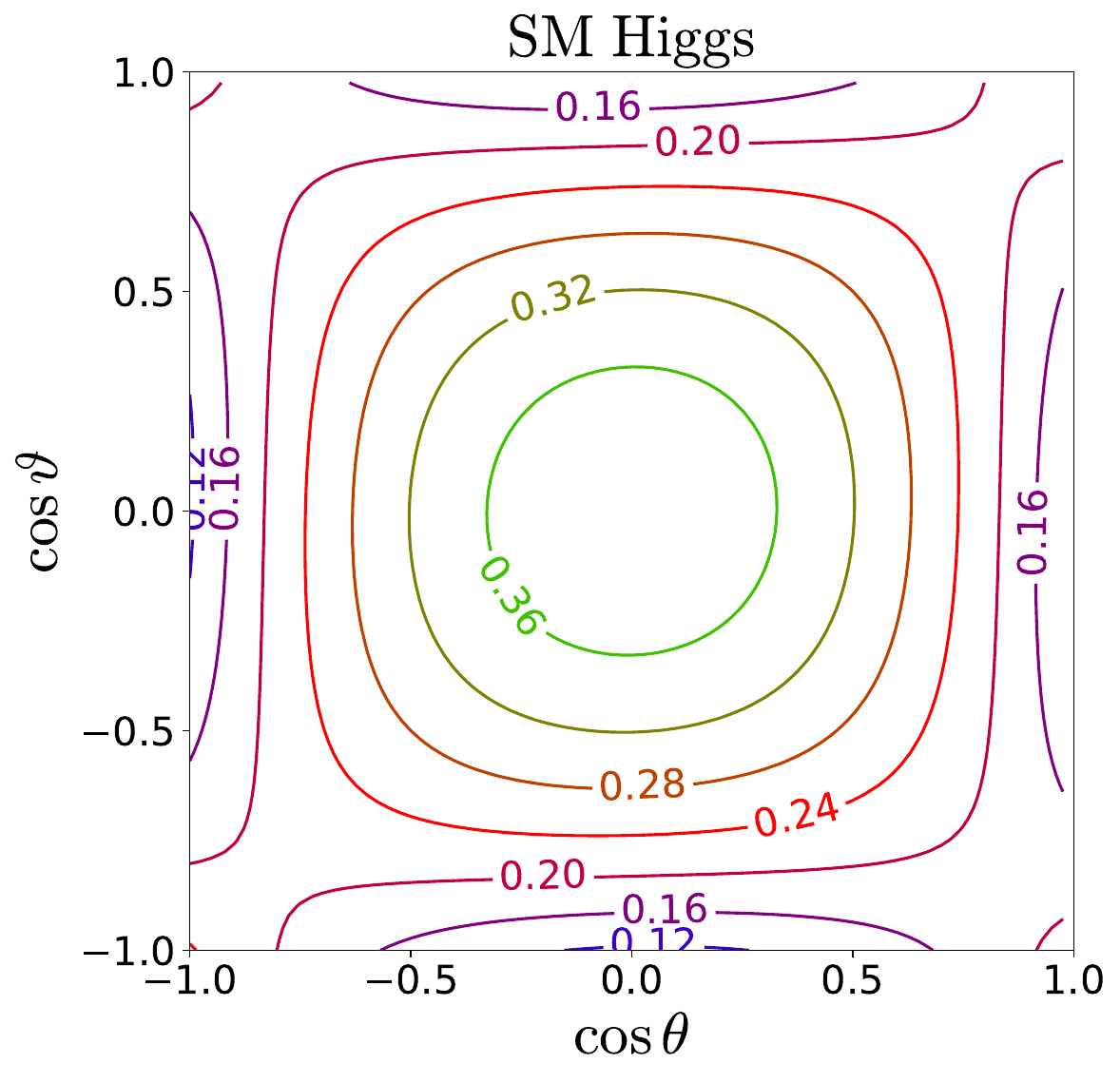}
\qquad\quad
\includegraphics[width=0.4\textwidth]{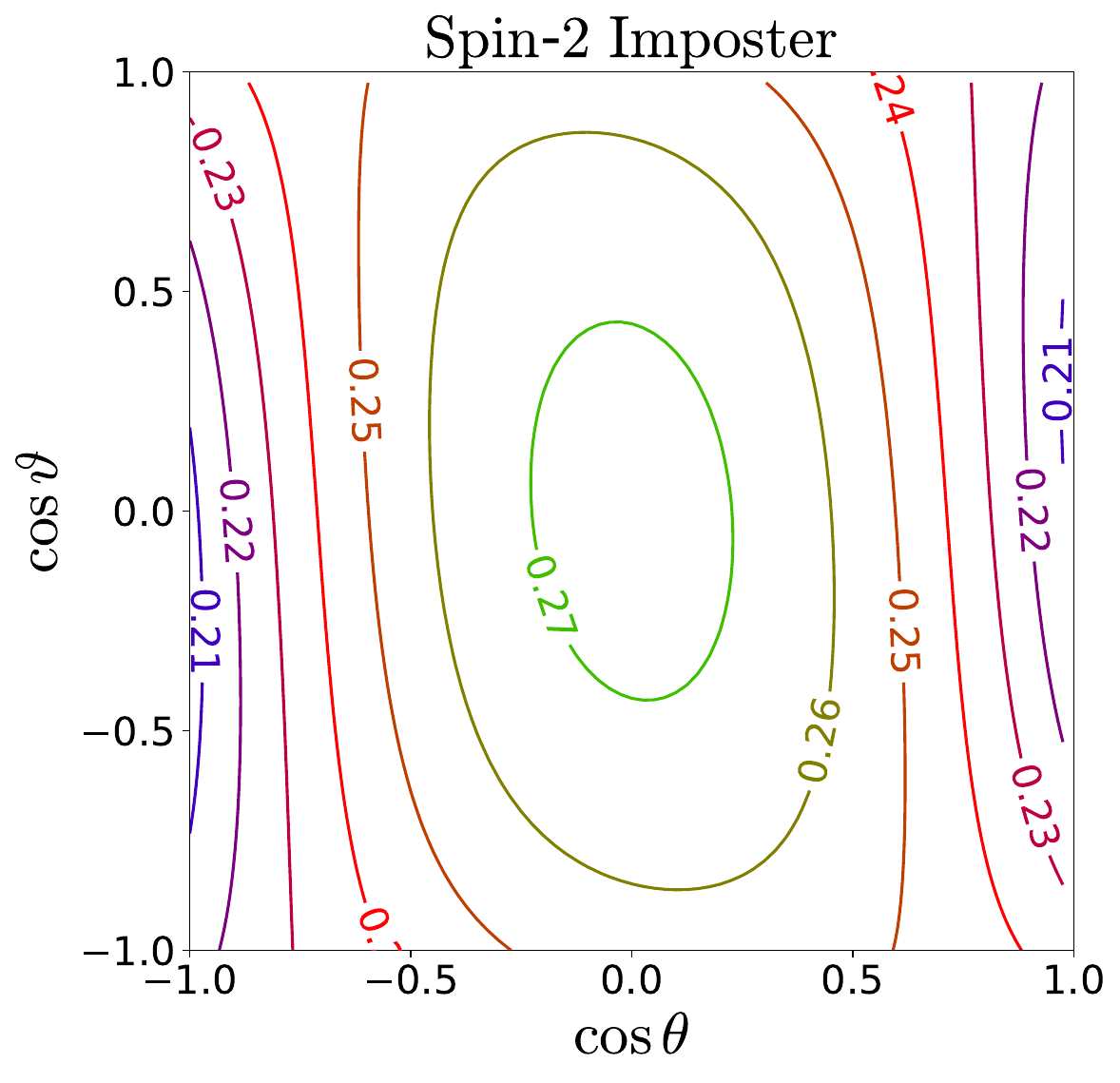}
\end{center}
\caption{Fully-correlated polar-angle distribution for 
        the SM Higgs boson (left) and the spin-2 KK-graviton imposter (right),
        respectively. As before, the $Z$-boson and $H$-boson masses are set 
        to be $m_Z=91\,{\rm GeV}$ and $m_H=125\, {\rm GeV}$, respectively. 
        }
\label{fig:double_polar_angle_distribution}
\end{figure}

One of the powerful quantities for vetoing the spin-2 KK graviton 
imposter against
the SM Higgs boson is the polar-angle correlation. As illustrated in 
Fig.$\,$\ref{fig:double_polar_angle_distribution}, 
the double polar-angle correlation 
for the SM Higgs boson (left frame) is clearly distinct from that for the
spin-2 KK graviton imposter, enabling us to veto the spin-2 imposter 
unambiguously. The slightly tilted distribution is due to the 
$\eta_\ell\eta_f \cos\theta\cos\vartheta$ originating from the 
parity-violating $Z^\star \ell\ell$ and $Z f\bar{f}$ couplings.

\subsection{The case with contact terms of maximal $j_0$}
\label{subsec:with_maximal_j0_contact_terms}

Generally, the four-point $H\ell\ell Z$ interactions contain not only
$j_0=1$ terms but also $j_0> 1$ terms for the $H$ spin $s_H\geq 1$, 
arising from new physics 
beyond the SM. They can be generated by simple 4-point 
angle-independent contact terms 
or terms describing the interactions of $\ell^\mp$ directly either 
with the $H$ boson or the on-shell $Z$ boson.
As described before, apart from the purely angle-dependent
Wigner $d$ functions, the reduced helicity amplitudes 
$\mathcal{T}^{[H\ell\ell Z]}_{\lambda_H;
\sigma,\lambda_Z}(m_\star, \kappa\cos\theta)$ in 
Eq.~\eqref{eq:3-body_cc_decay_helicity_amplitude} 
is in general dependent on the combination $\kappa\cos\theta$
vanishing at the threshold with $m_\star=m_H-m_Z$. In contrast,
each contact term with maximal $j_0$ generates a constant reduced
helicity amplitude.\s

\begin{itemize}
\item  {\bf Spin 0.} If the $H$ particle is spinless, the $j_0$ value is fixed
to be 1. Therefore, the parity-even case with contact 
terms could imitate  the case with the SM Higgs boson for the polar and 
azimuthal angular distributions. However, with no virtual $Z^\star$ exchange, 
the overall pattern of the two-lepton invariant mass distribution is 
qualitatively different from the SM case  for which the distribution 
varies in proportion
to $\omega^4_Z/(\omega^2_Z-\omega^2_\star)^2$ as 
$\omega_\star$ approaches the value of $1-\omega_Z=0.27$ at the threshold.
For example, the contact term of the parity-even 
$U^+_{\mu\beta}+U^-_{\mu\beta}$ form leads to the normalized 
invariant-mass distribution as
\begin{eqnarray}
      \frac{d{\cal N}_H}{dm_\star}
\, =\,  \frac{m^3_\star}{m^4_H\, {\cal C}_1(\omega_Z)}\, \kappa
\quad \mbox{with}\quad 
{\cal C}_1(\omega_Z)=\omega^2_Z\, A_2(\omega_Z)
                    + A_3(\omega_Z),
\end{eqnarray}
when the theory is nearly $P$ invariant. The explicit form of 
${\cal C}_1(\omega_Z)$ is given
in Appendix~\ref{appedix:integral_functions}. The single polar-angle 
distribution due to the contact term is of the simple form
\begin{eqnarray}
     \frac{d{\cal N}_H}{d\cos\theta}
\,=\, \frac{3}{8} (1+\cos^2\theta),
\end{eqnarray}
clearly different from that of the SM Higgs boson.
On the other hand, the parity-odd pseudoscalar case can be
ruled out unambiguously by confirming not only the invariant mass but also
and the polar and azimuthal angle correlations characterizing the SM Higgs 
boson, as before.
\item {\bf Spin 1.} If the $H$ spin is 1, the general spin-1 four-point 
vertex tensor contains two non-trivial $j_0=2$ contact vertex terms as
\begin{eqnarray}
\omega_\star W^{\mu_1}_\sigma \hat{l}^{\mu_2}
U^\pm_{\alpha\mu_1}U^\pm_{\beta\mu_2},
\label{spin-1_contact_term}
\end{eqnarray}
which correspond to the $H$ and $Z$ helicity combinations, 
$(\pm 2, \mp 1)$, generating the distributions of the maximal 
$j_0$ value of 2. Because each of them contains two terms 
with no momentum dependence, the corresponding normalized invariant-mass distribution decreases steeply as
\begin{eqnarray}     
\frac{d{\cal N}_H}{dm_\star}
\, =\,  \frac{m^5_\star}{m^4_H\, {\cal C}_2(\omega_Z)}\, \kappa,
\quad \mbox{with}\quad 
{\cal C}_2(\omega_Z)=\omega^4_Z\, A_2(\omega_Z)+2\omega^2_Z A_3(\omega_Z)
                    + A_4(\omega_Z),
\end{eqnarray}
which is linearly proportional to $\kappa$ near the threshold.  
For the explicit form of ${\cal C}_2(\omega_Z)$ we refer to 
Appendix~\ref{appedix:integral_functions}.
Nevertheless, due to the absence of any intermediate state, 
the global feature of the invariant mass distribution will be 
qualitatively different from the SM case.\s

Furthermore, combined with 
the $\ell^-\ell^+$ system, the contact terms in 
Eq.~\eqref{spin-1_contact_term} generate 
the non-trivial Wigner $d$ functions 
\begin{eqnarray}
  d^2_{\pm 2, \sigma} (\theta)
= -\frac{1}{2} \sin\theta\,(1\pm\sigma\cos\theta)\ \
\mbox{with} \ \ \sigma=\pm 1,
\end{eqnarray}
as shown in Eq.$\,$\eqref{eq:l-dependent_d_functions}.
Therefore, the single polar-angle distribution due to the $j_0=2$ 
contact terms in the spin-1 case is given by
\begin{eqnarray}
  \frac{d{\cal N}_H}{d\cos\theta}
= \frac{5}{8}\sin^2\theta\, (1+\cos^2\theta),
\end{eqnarray}
which is distinctly different from that of the SM Higgs boson.
\item {\bf Spin 2.} If the $H$ spin is 2, we have eight ($8=2\times 4$)
$j_0=2$ terms and there are four ($4=2\times 2$) $j_0=3$ terms as shown
in Tab.$\,$\ref{tab:spin-2_covariant_operators}. Every $j_0=2$ term 
involves at least one power of momentum so that every helicity amplitude
vanishes near threshold linearly in $\kappa$, distinct from the SM.
On the other hand, each of the $HZ$ operators with $j_0=3$ contains 
a momentum-free term leading to non-zero helicity amplitudes at 
the threshold. The contact terms consisting of the $HZ$ operators with 
$j_0=3$ generate the following normalized invariant-mass distribution:
\begin{eqnarray}     
\frac{d{\cal N}_H}{dm_\star}
\! =\!  \frac{m^7_\star}{m^8_H  {\cal C}_3(\omega_Z)} \kappa,
\end{eqnarray}
where the $\omega_Z$-dependent integral function is given by
\begin{eqnarray}
  {\cal C}_3(\omega_Z)
\, =\, \omega^6_Z\, A_2(\omega_Z) + 3 \omega^4_Z A_3(\omega_Z)
                   + 3\omega^2_Z A_4(\omega_Z)
                    + A_5(\omega_Z),
\end{eqnarray}
of which the explicit form is 
given in Appendix~\ref{appedix:integral_functions}.
The normalized invariant-mass distribution decreases steeply 
near the threshold.\s

Combined with the $\ell^-\ell^+$ system,
the terms generate the helicity amplitudes proportional to 
the non-trivial Wigner $d$ functions
\begin{eqnarray}
  d^3_{\pm 3, \sigma} (\theta)
= \frac{\sqrt{15}}{8}\sin^2\theta\, (1\pm \sigma\cos\theta)
\ \ \mbox{with} \ \ \sigma=\pm 1.
\end{eqnarray}
Therefore, the normalized single polar-angle distribution 
due to the $j_0=s_H+1=3$ contact term in the spin-2 case  is given by
\begin{eqnarray}
  \frac{d{\cal N}_H}{d\cos\theta}
= \frac{105}{128}\sin^4\theta\, (1+\cos^2\theta),
\end{eqnarray}
which is distinctly different from that of the SM Higgs boson.\s

Figure~\ref{fig:numerical_examples_with_contact_terms} shows 
the invariant-mass distribution on the left frame and the single polar-angle 
distribution on the right frame of the cases only with the $s_H=0$ contact 
(red dashed line), $s_H=1$ contact (blue dot-dashed line), and  $s_H=2$ contact
(green dotted line) terms, each of which corresponds to the 
maximally-allowed $j_0$ value of 1, 2, or 3, respectively, in comparison  
with the SM Higgs boson 
(black solid line). Even in the spin-0 case, the invariant-mass distribution
is different from that of the SM Higgs boson, with more prominent difference
for the higher-spin case, as shown in the left frame of
Fig.~\ref{fig:numerical_examples_with_contact_terms}. The single polar-angle
distribution due to any maximal-$j_0$ contact term is also very different
from that of the SM Higgs boson as clearly shown in the right frame of
Fig.~\ref{fig:numerical_examples_with_contact_terms}.

\begin{figure}[htb!]
\vskip 0.4cm
\begin{center}
\includegraphics[width=0.42\textwidth]{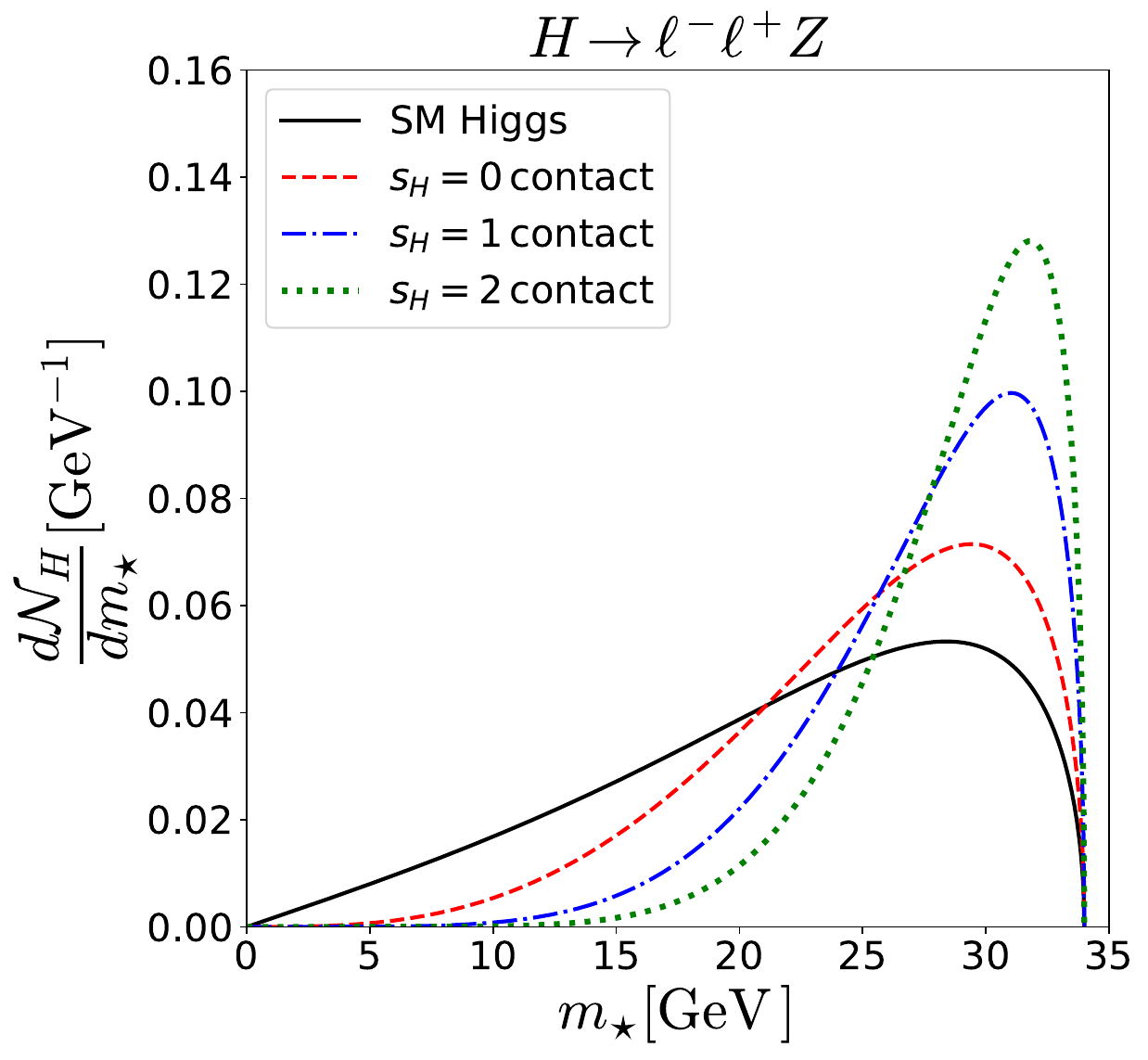}
\qquad\quad
\includegraphics[width=0.42\textwidth]{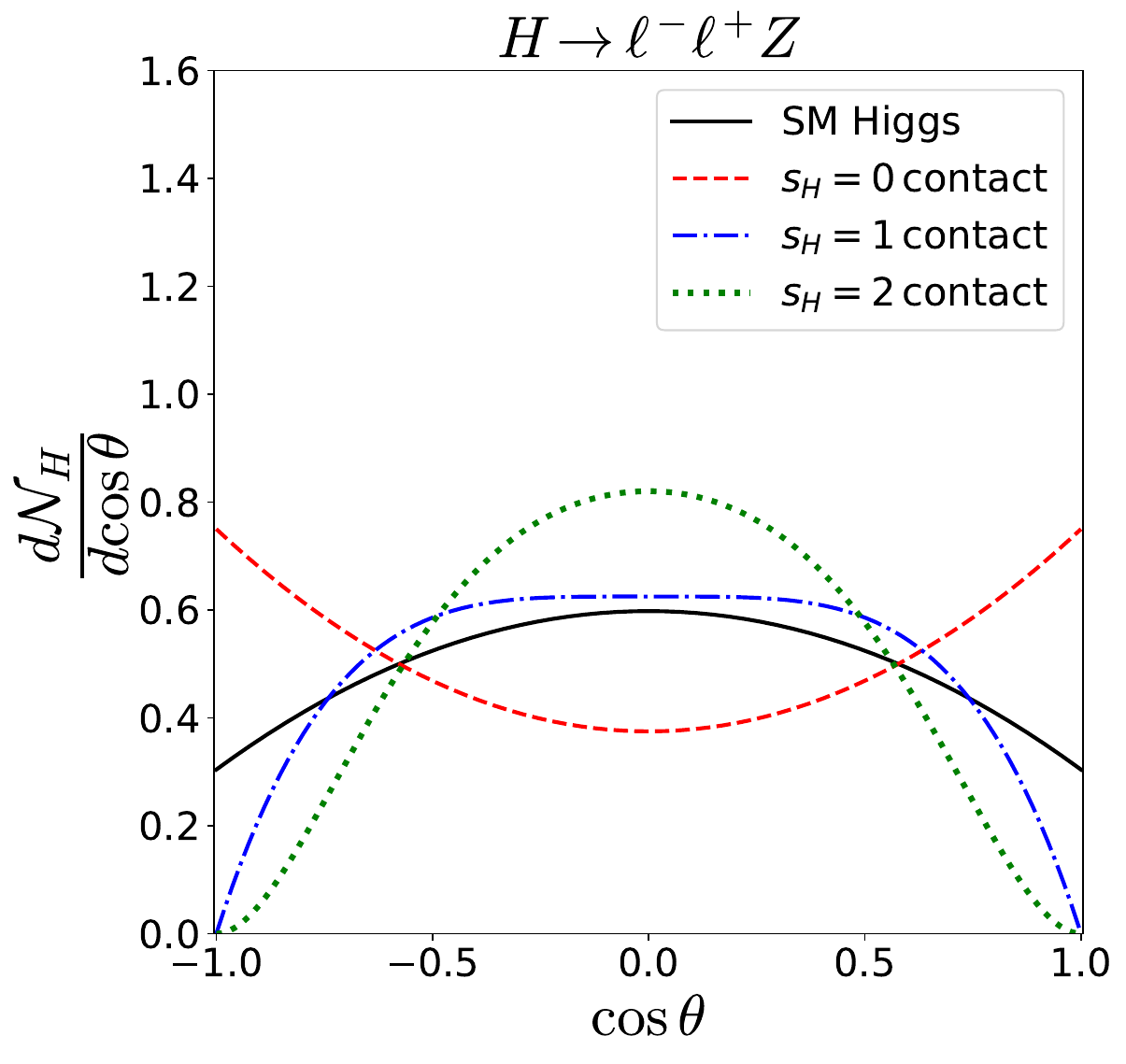}
\end{center}
\caption{Invariant-mass (left frame) and single polar-angle (right frame)
    distributions of the cases only with the $s_H=0$ contact  
    (red dashed line), $s_H=1$ contact (blue dot-dashed line), 
    and $s_H=2$ contact (green dotted line) terms corresponding 
    to the maximal $j_0$ values of 1, 2, and 3, respectively, in comparison 
    with the SM Higgs boson (black solid line). Both distributions
    due to the maximal-$j_0$ contact terms in any integer-spin case are 
    clearly distinct from those of the SM Higgs boson.}
\label{fig:numerical_examples_with_contact_terms}
\end{figure}
\item {\bf Spin \boldmath{$\geq 3$}.} Except for four special cases to be
mentioned shortly, every term involves at least one power of momentum
so that the invariant mass distribution vanishes near the threshold linearly
at least in $\kappa^3$, distinct from the SM case. The four $HZ$ operators
with no momentum dependence  are
\begin{eqnarray}
&&    [{\cal H}^{(s_H+1)[HZ]}_{\pm s_H, \pm 1}]
 \, =\,  [U^\pm_H]^{s_H} [U^\pm_Z], \\
&&    [{\cal H}^{(s_H-1)[HZ]}_{\pm s_H, \mp 1}]
 \, =\, [U^\pm_H]^{s_H-1} [U^\pm_{HZ}].
\end{eqnarray}
Combining these $HZ$ and $\ell\ell$ operators with all the wave functions
generates the helicity amplitudes with the following 
non-trivial Wigner $d$ functions (apart from some $\kappa\cos\theta$-dependent
form factors)
\begin{eqnarray}
             d^{(s_H+1)}_{\pm (s_H+1), \sigma} (\theta)
& \propto & (\sin\theta)^{s_H}\, (1\pm \sigma\cos\theta),\\
             d^{(s_H-1)}_{\pm (s_H-1), \sigma} (\theta)
& \propto & (\sin\theta)^{s_H-2}(1\pm \sigma\cos\theta).
\end{eqnarray}
Clearly, for $s_H\geq 3$, the non-trivial power terms of $\sin\theta$ render
the single polar-angle distribution significantly different from that of 
the SM Higgs boson with the spin value of $s_H=0$. Explicitly,
the normalized invariant-mass distribution for the maximal $j_0=s_H+1$ 
term is given by
\begin{eqnarray}      
   \frac{d{\cal N}_H}{dm_\star}
=  \frac{m^{2s_H+3}_\star}{m^{2(s_H+2)}_H \mathcal{C}_n(\omega_Z)}\, \kappa,
\end{eqnarray}
where the $\omega_Z$-dependent integral function ${\cal C}_n(\omega_Z)$ is
defined in general by
\begin{eqnarray}
  \mathcal{C}_n(\omega_Z) 
= \int^{1-\omega_Z}_{0} d\omega_\star
               \omega^{2n+1}_\star\, \kappa,
\end{eqnarray}
in terms of the non-negative integer $n$, 
which can be expressed as a linear combination of the $A$ integral
functions defined in Eq.$\,$\eqref{eq:a_integral_functions}.\s

The maximal-$j_0$ term with $j_0=s_H+1$ leads to the corresponding normalized 
single polar-angle distribution as
\begin{eqnarray}
  \frac{d{\cal N}_H}{d\cos\theta}
= \frac{(2s_H+3)}{2^{2s_H+2}(s_H+2)}\,
   \frac{(2s_H+1)!}{(s_H!)^2} \, (\sin\theta)^{2s_H}\, (1+\cos^2\theta),
\end{eqnarray}
valid for any non-negative integer
spin $s_H$ including $s_H=0, 1$ and $2$ and manifestly different from the
SM Higgs case for positive values of $s_H$.
\end{itemize}

\setcounter{equation}{0}

\section{Summary and conclusion}
\label{sec:summary_conclusion}

We have devised an effective and systematic algorithm for constructing covariant 
four-point $H\ell\ell Z$ vertices pertaining to a massive particle $H$ with 
an arbitrary integer spin value. This algorithm leverages a direct correspondence
between helicity formalism and covariant representation to delineate the 3-body 
decay channel $H\to \ell^-\ell^+ Z$. By upholding chirality conservation 
and disregarding the negligible masses of light leptons (where $m_\ell=0$ for 
$\ell=e, \mu$), our algorithm stands as a natural yet intriguing extension 
of the previously established methodology for systematically constructing 
all-encompassing three-point vertices detailed across a series of
publications~\cite{Choi:2021ewa,Choi:2021qsb,Choi:2021szj}.\s

The methodically developed general covariant four-point $H\ell\ell Z$ 
vertices serve as a robust tool for systematically discerning the spin 
and parity attributes of the SM Higgs boson, particularly within the 
realm of one of its key decay channels, the three-body decay 
$H\to \ell^-\ell^+ Z$, where $\ell=e, \mu$,  observable at the LHC. 
This analytical approach, centered around the general covariant 
four-point vertex, surpasses preceding studies on Higgs spin and 
parity determination. Furthermore it accommodates a much broader 
spectrum of analyses, encompassing all the effectively allowed 
4-point $H\ell\ell Z$ interactions.\s

We have confirmed firmly that threshold effects and angular correlations 
can be adopted to determine spin and parity unambiguously  in the significantly
expanded setting, though high 
event rates are required for the analysis in practice. These essential
conclusions can be covered by summarizing the analyses in a few 
characteristic points. The SM Higgs boson possesses a unique 
invariant-mass spectrum with its steeply-decreasing threshold behavior 
and also its unique polar-angle and azimuthal-angle
correlations, characterizing its spin-zero and parity-even nature and
its messenger role of EWSB. With the observation of the unique global 
spectrum and the steep decrease near the kinematical limit for the 
SM Higgs boson, we can rule out all the other spin and/or parity assignments 
to the $H$ boson not only in the standard two-step cascade decay but also 
in all the decay processes involving the general four-point interactions
by checking out the two-lepton invariant-mass distribution and  the single and/or
correlated polar-angle and azimuthal-angle distributions.\s

The selection rules proposed so far for confirming the spinless nature and
even parity of the SM Higgs boson unambiguously through its 3-body decay 
$H\to \ell^-\ell^+Z$  can be supplemented by observations specific to two cases. 
By observing non-zero $H\gamma\gamma$ and $Hgg$ couplings, the unit-spin 
assignment of $s_H=1$ can elegantly ruled out in particular and 
every odd-spin assignment in general by Landau-Yang 
theorem~\cite{Landau:1948kw,Yang:1950rg,Choi:2012yg,
Ellis:2012wg,CMS:2014afl}.\s

The formalism developed in the present work can be extended to exclude mixed 
normality states effortlessly. The differential decay 
width in the $CP$-noninvariant theory proves to be more intricate, compared to 
that in the $CP$-invariant theory. Particularly, the double polar-angle distribution 
is adjusted to incorporate linear terms that are proportional to either $\cos\theta$
or $\cos\vartheta$, demonstrating $CP$ violation~\cite{Kramer:1993jn,
Hagiwara:1993sw,Hagiwara:2000tk,Grzadkowski:2000hm,Han:2000mi}. 
Despite the inherent complexity, the analysis for determining the spin of the 
SM Higgs particle by ruling out all potential imposter scenarios follows 
the same procedure as in the case of fixed normality, since the general 
4-point vertex is simply the sum of even and odd normality tensors.\s

The general algorithm for constructing the 
covariant four-point $H\ell\ell Z$ vertex with $H$ of any integer 
spin enables us to work out a similar unambiguous and powerful procedure 
for identifying the spin and parity of the SM Higgs boson in 
the so-called Higgs-strahlug process 
$e^-e^+\to ZH$, topologically equivalent to the 3-body decay $H\to e^-e^+Z$. 
This investigation is currently under rigorous examination and is 
anticipated to reach completion soon.\s 

Absolutely, even after confirming the spinless nature of the $H$ boson,  
the general Lorentz-covariant four-point vertex structure serves as 
a robust foundation not only for delving deeper into the 3-body decay 
process of the spin-0 Higgs boson but also for exploring a range of 
theoretical considerations such as unitarity and potential (hidden) 
discrete and/or gauge symmetries through diverse two-to-two scattering 
and three-body decay processes~\cite{Cornwall:1973tb,Cornwall:1974km,
LlewellynSmith:1973yud,Lee:1977yc,Lee:1977eg}.\s

Before closing, we mention that it will be extremely valuable to develop a 
computation package implementing our general analytic algorithm fully 
so as to identify the
spin and parity of the SM Higgs boson and to investigate various new physics
phenomena involving any integer-spin massive particle automatically with 
real experimental data delivered at not only the LHC but also 
the high-luminosity
LHC (HL-LHC)~\cite{Mangano:2019kji,Shiltsev:2019rfl,Rossi:2015ulc}.\s

\vskip 1.0cm

\section*{Acknowledgments}

This work is supported by the Basic Research Laboratory Program of
the National Research Foundation of Korea
(Grant No. NRF-2022R1A4A5030362 for  SYC and DWK).
SYC is supported in part by the Basic Science Research Program of
Ministry of Education through the National Research Foundation of
Korea (Grant No. NRF-2022R1I1A3071226). JJ is supported by a KIAS
Individual Grant (QP090001) via the Quantum Universe Center
at Korea Institute for Advanced Study.\s

\vskip 1.0cm

\appendix

\setcounter{equation}{0}

\section{Various formulas for evaluating helicity amplitudes}
\label{appendix:analytic_formulas}

It is necessary to derive the helicity amplitudes 
(\ref{eq:3-body_cc_decay_helicity_amplitude}) for the three-body decay
process $H\to\ell^-\ell^+ Z$ by calculating the corresponding amplitudes
(\ref{eq:amplitudes_with_covariant_operators}) expressed in terms
of the general $HZ$ and $\ell\ell$ operators before investigating
all the invariant-mass and angular correlations systematically.
We collect in this appendix~\ref{appendix:analytic_formulas} a set of
analytic formulas which can be exploited for calculating the helicity 
amplitudes systematically and efficiently. \s

Firstly, for the sake of self-containment and concreteness, we re-write down 
all the relevant normalized momenta introduced in the main text
collectively:
\begin{eqnarray}
\hat{p} &=& (1,0,0,0) \quad\, \Rightarrow \quad 
            \hat{p}_{\star\mu} = 2\omega_\star\hat{p}_\mu,\quad
            \hat{p}_{Z\beta} = 2\omega_Z \hat{p}_\beta,\\
\hat{r} &=& (0,0,0, 1) \quad\ \ \Rightarrow\quad 
            \hat{r}_{H\alpha} = \kappa\hat{r}_\alpha,\\
\hat{k} &=& \frac{1}{2\omega_\star} (e_\star, 0,0 ,\kappa)\ \ 
            \mbox{and}\ \ 
\hat{l}  = \frac{1}{2\omega_\star} (\kappa\cos\theta,\, 
            2\omega_\star\sin\theta,
             0, e_\star\cos\theta), 
\end{eqnarray}
along with the normalized chirality-conserving $\ell^-\ell^+$ lepton 
current $\hat{l}_\pm$
\begin{eqnarray}
\hat{l}_\pm = \pm \frac{1}{2\sqrt{2}\omega_\star}
            (\kappa\sin\theta, -2\omega_\star\cos\theta,\,
             \pm 2i\omega_\star,
             e_\star\sin\theta),
\end{eqnarray}
in the $H$ rest frame. Here, we introduce the normalized masses, $\omega_\star=m_\star/m_H$ and 
$\omega_Z=m_Z/m_H$, defining
the kinematic variables, $e_Z=1+\omega^2_Z-\omega^2_\star$ and
$e_\star=1-\omega^2_Z+\omega^2_\star$, and the
polar angle $\theta$ of the flight direction of the lepton $\ell^-$  
in the $\ell^-\ell^+$ rest frame with respect to the 
original 3-momentum of the $\ell^-\ell^+$ system in the kinematic 
configuration in Fig.$\,$\ref{fig:h_to_llz_rm} where 
the $H$ boson is at rest. Contracting the newly-defined momenta and their 
corresponding longitudinal polarization vectors or normalized momenta 
lead to the following set of non-trivial relations
\begin{eqnarray}
    \hat{r}_{H}\cdot \epsilon(p_H,\lambda_H)
= -\kappa\, \delta_{\lambda_H, 0} \ \ &\mbox{and}& \ \
   \epsilon^{*}(k_Z, \lambda_Z)\cdot \hat{p}_{Z} 
 = \kappa\, \delta_{\lambda_Z, 0},\\
   \hat{p}_{\star}\cdot \hat{l}_{\pm} 
= \pm \frac{1}{\sqrt{2}} \kappa \sin\theta \ \ &\mbox{and}& \ \
   \hat{p}_{\star}\cdot \hat{l}
= \kappa \cos\theta,
\end{eqnarray}
where the longitudinal $H$ and $Z$ polarization vectors are explicitly given by
\begin{eqnarray}
\epsilon(p_H, 0) = (0,0,0,1)\quad \mbox{and}\quad
\epsilon(k_Z, 0) = \frac{1}{2\omega_Z}(\kappa, 0,0, -e_Z).
\end{eqnarray}
It is worthwhile to note that each contracted term is linearly proportional to 
the kinemtical factor $\kappa$,
becoming zero at the kinematical threshold of 
$\omega_\star=1-\omega_Z$.\s

Secondly, in the same kinematic configuration of 
Fig.$\,$\ref{fig:h_to_llz_rm}, the basic operator $U^\pm$ in any 
pair of three 4-vector indices $\{\alpha, \mu, \beta\}$, is given 
in a $4\times 4$ matrix form by
\begin{eqnarray}
U^\pm = -\frac{1}{2} 
       \left(\begin{array}{cccc}
             0 &   0   &   0   & 0 \\
             0 &   1   & \mp i & 0 \\
             0 & \pm i &  1    & 0 \\
             0 &   0   &  0    & 0 
             \end{array}
       \right) 
      = -|p_H,\pm 1\rangle \langle p_H,\pm 1|
      = -|k_Z,\mp 1\rangle \langle k_Z,\mp 1| ,
\end{eqnarray}
which is corresponding to the tensor product of the transverse 
polarization 4-vector,  $\epsilon(p_H,\pm 1)$ or $\epsilon(k_Z,\mp 1)$, 
of which the expression is given by
\begin{eqnarray}
 \epsilon(p_H,\pm 1)
 = \epsilon^*(k_Z,\pm 1)= \frac{1}{\sqrt{2}}(0, \mp 1, -i, 0),   
\end{eqnarray}
respectively, and its Hermitian conjugate vector,  $\epsilon^*(p_H,\pm 1)$
or $\epsilon^*(k_Z,\mp 1)$. Contracting the tensor 
operator with any pair of the transverse $H$ and $Z$ polarization vectors, 
the normalized lepton current $\hat{l}_\sigma$ and the normalized momentum 
$\hat{l}$ leads to the following set of non-trivial
relations:
\begin{eqnarray}
&& \epsilon^{*}(k_Z,\lambda_Z)\cdot 
   U^\pm\cdot \epsilon(p_H,\lambda_H)
  = -\delta_{\lambda_H,\pm 1}\, \delta_{\lambda_Z, \mp 1}, 
  \label{eq:contraciton_u_ba} \\
&& \hat{l}_\sigma\cdot  
   U^\pm\cdot  \epsilon(p_H,\lambda_H)
   \,\, =\, \mp \frac{1}{2}(1\pm\sigma\cos\theta)\, \delta_{\lambda_H,\pm1}, 
   \label{eq:contraciton_u_ma_l_trans} \\ 
&&  \hat{l}\cdot  
   U^\pm\cdot  \epsilon(p_H,\lambda_H)
    \,\,\,\, =\,\,\, \frac{1}{\sqrt{2}}\sin\theta\, \delta_{\lambda_H, \pm 1}, 
    \label{eq:contraciton_u_ma_l_longi} \\
&&  \hat{l}_\sigma  \cdot 
   U^\pm\cdot \epsilon^{*}(k_Z,\lambda_Z)
   =\, \mp\frac{1}{2}(1\pm\sigma\cos\theta)\, \delta_{\lambda_Z,\mp 1}, 
   \label{eq:contraciton_u_mb_l_trans}\\
&& \hat{l} \cdot 
   U^\pm\cdot \epsilon^{*}(k_Z,\lambda_Z)
   \,\,\, =\, \frac{1}{\sqrt{2}}\sin\theta\, \delta_{\lambda_Z, \mp 1},  
   \label{eq:contraciton_u_mb_l_longi}
\end{eqnarray}
of which all terms depend only on the polar angle $\theta$. \s

\setcounter{equation}{0}

\section{Integral functions}
\label{appedix:integral_functions}

In order to facilitate the integration of various $\omega_\star$-dependent 
functions and to express the results in a compact form, we 
introduce two types of $\omega_Z$-dependent integral functions, $A_n$ and $B_n$, 
as
\begin{eqnarray}
     A_n(\omega_Z)
&=& \int^{1-\omega_Z}_0 d\omega_\star\,\,
    \frac{\omega_\star\, \kappa}{(\omega_\star^2-\omega_Z^2)^{2-n}},
\label{eq:a_integral_functions}
    \\
     B_n(\omega_Z)
&=& \int^{1-\omega_Z}_0 d\omega_\star\,\,
    \frac{\omega^2_\star\, \kappa}{(\omega_\star^2-\omega_Z^2)^{2-n}},
\label{eq:b_integral_functions}
\end{eqnarray}
for a non-negative integer $n$. \s

Firstly, the  expressions of six integral functions $A_n$ from $n=0$ to $n=5$ 
are given explicitly by
\begin{eqnarray}
    A_0(\omega_Z)
&=& -\frac{1-\omega_Z^2}{2\omega_Z^2}
   +\frac{1}{2\sqrt{4\omega_Z^2-1}}\cos^{-1}
   \bigg(\frac{3\omega_Z^2-1}{2\omega_Z^3}\bigg)
   -\frac12 \ln \omega_Z,\\
   A_1(\omega_Z)
&=& -\frac{1-\omega_Z^2}{2}
    +\frac{\sqrt{4\omega_Z^2-1}}{2}\cos^{-1}
    \bigg(\frac{3\omega_Z^2-1}{2\omega_Z^3}\bigg)
    +\frac{1}{2} \ln \omega_Z,\\
   A_2(\omega_Z)
&=& \frac{1-\omega_Z^2}{4}(1+\omega_Z^2)
    +\omega_Z^2 \ln \omega_Z,\\
    A_3(\omega_Z)
&=& -\frac{1-\omega_Z^2}{12}
    (2\omega_Z^4-7\omega_Z^2-1)
    +\omega_Z^2 \ln \omega_Z,\\
    A_4(\omega_Z)
&=& \frac{1-\omega_Z^2}{24}
    (3\omega_Z^6-5\omega_Z^4+25\omega_Z^2+1)
    +\omega_Z^2(1+\omega_Z^2) \ln \omega_Z,\\
    A_5(\omega_Z)
&=& -\frac{1-\omega_Z^2}{120}
    (12\omega_Z^8-13\omega_Z^6-73\omega_Z^4-163\omega_Z^2-3)
    +\omega_Z^2(1+3\omega_Z^2) \ln \omega_Z,
\end{eqnarray}
with $\omega_Z=m_Z/m_H=0.73$ less than unity.\s

Secondly, the expressions of five integral functions  $B_n$ from $n=0$ 
to $n=4$ are given 
explicitly in terms of three types of elliptic integrals by
\begin{eqnarray}
   B_0(\omega_Z)
&=& \frac{2\omega_Z^2+6\omega_Z-1}{2(1+\omega_Z)}\, 
     K\left( r_Z\right)
    -\frac{3(1+\omega_Z)}{2}\, E\left( r_Z\right) 
    +\frac{6\omega_Z^2-1}{2\omega_Z^2(1+\omega_Z)} 
     \Pi\left(n_Z, r_Z\right),\\
   B_1(\omega_Z)
&=& \frac{\omega_Z(2\omega_Z^2-11\omega_Z-4)}{3(1+\omega_Z)}\, 
     K\left( r_Z\right)
   + \frac{(1+\omega_Z)(2-\omega_Z^2)}{3} \, 
     E\left( r_Z\right) \nonumber\\
&& + \frac{4\omega_Z^2-1}{1+\omega_Z}\,  
     \Pi\left(n_Z, r_Z\right),\\
   B_2(\omega_Z)
&=& -\frac{2(1+\omega_Z)}{15}
\Big[2\omega_Z(\omega_Z^2+6\omega_Z+1)
     K\left( r_Z\right)-(\omega_Z^4+14\omega_Z^2+1)\, 
     E\left( r_Z\right)\Big],\\
  B_3(\omega_Z)
&=& \frac{2(1+\omega_Z)}{105}
\Big[2\omega_Z(4\omega_Z^4-6\omega_Z^3-19\omega_Z^2-48\omega_Z-3)
     K\left( r_Z\right)
     \nonumber\\
&&  -(4\omega_Z^6-27\omega_Z^4-118\omega_Z^2-3)\, 
     E\left( r_Z\right)\Big],\\
  B_4(\omega_Z)
&=& -\frac{2(1+\omega_Z)}{315}
\Big[2\omega_Z(8\omega_Z^6-12\omega_Z^5-12\omega_Z^4
+180\omega_Z^3+105\omega_Z^2+150\omega_Z+5)
     K\left( r_Z\right)
     \nonumber\\
&&  -(8\omega_Z^8-28\omega_Z^6+453\omega_Z^4+410 \omega_Z^2+5)\, 
     E\left( r_Z\right)\Big],
\end{eqnarray}
with $n_Z=(1-\omega_Z)/\omega_Z$, and $r_Z=(1-\omega_Z)/(1+\omega_Z)$.
The first-, second- and third-kind complete elliptic integrals are defined 
by 
\begin{eqnarray}
   K\left(r_Z\right)
&=& \int^{\pi/2}_{0} \frac{d\theta}{\sqrt{1-r^2_Z\sin^2\theta}}, \\
   E\left(r_Z\right)
&=& \int^{\pi/2}_{0} \sqrt{1-r^2_Z\sin^2\theta}\, d\theta, \\
    \Pi\left(n_Z , r_Z \right)
&=& \int^{\pi/2}_{0}  
    \frac{1}{1-n_Z^2\sin^2\theta}
    \frac{d\theta}{\sqrt{1-r_Z^2\sin^2\theta}},
\end{eqnarray}
in a definite integral form.\s

The invariant-mass and angular distributions for the pure 
contact interactions  without any intermediate virtual $Z$ exchange 
can be expressed in a compact manner by introducing 
another set of $\omega_Z$-dependent integral functions as
\begin{eqnarray}
\mathcal{C}_n(\omega_Z) = \int^{1-\omega_Z}_{0} d\omega_\star\,
               \omega^{2n+1}_\star\, \kappa.
\end{eqnarray}
Explicitly, the expressions of the integrals, ${\cal C}_n(\omega_Z)$, 
for $n=1,2,3$ are given by,
\begin{eqnarray}
   \mathcal{C}_1(\omega_Z)
&=& \frac{1-\omega_Z^4}{4}
    +\omega_Z^2\ln \omega_Z,\\
  \mathcal{C}_2(\omega_Z)
&=& \frac{1-\omega_Z^2}{12}(1+10\omega_Z^2 + \omega_Z^4)
    +\omega_Z^2(1+\omega_Z^2) \ln \omega_Z,\\
    \mathcal{C}_3(\omega_Z)
&=& \frac{1-\omega_Z^2}{24}
    (1 + 29  \omega_Z^2 + 29 \omega_Z^4 + \omega_Z^6)
    +\omega_Z^2(1+ 3\omega_Z^2 + \omega_Z^4) \ln \omega_Z.
\end{eqnarray}
As shown explicitly  with ${\cal C}_{1,2,3}$ in the main text, 
every ${\cal C}$ integral function can be
expressed as a linear combination of the $A$ integral functions defined in
Eq.$\,$\eqref{eq:a_integral_functions}.\s

\skip 1.0cm

\bibliographystyle{unsrt}
\bibliography{refs}

\end{document}